\documentclass{WileyMSP-template}

\usepackage{color}
\usepackage{amsmath,mathrsfs,stmaryrd,titletoc,subfigure}
\usepackage{amssymb,amsfonts,amsthm,stmaryrd}
\usepackage{upgreek}
\usepackage{setspace} 
\usepackage[numbers,super,square,sort&compress]{natbib}
\makeatletter
\renewcommand{\NAT@sep}{,}
\makeatother
\PassOptionsToPackage{colorlinks=true, linkcolor=blue, citecolor=blue, urlcolor=blue}{hyperref}
\usepackage{hyperref} % Load only once and last

\usepackage{tabularx} % complex tabular
\usepackage{booktabs} % complex tabular

\newcommand{\bfx}{\boldsymbol{x}}
\newcommand{\bfJ}{\boldsymbol{J}}
\newcommand{\bfu}{\boldsymbol{u}}
\newcommand{\bfsigma}{\boldsymbol{\sigma}}
\newcommand{\bfepsilon}{\boldsymbol{\epsilon}}

\newcommand{\bfn}{\boldsymbol{n}}
\newcommand{\bfi}{\boldsymbol{i}}

\begin{document}
	
	\pagestyle{fancy}
	\rhead{\includegraphics[width=2.5cm]{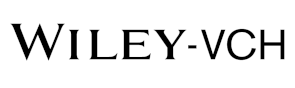}}
	
	\begin{spacing}{1.0}
		
		{\RaggedRight
			% \title{Deciphering the interplay between wetting and chemo-mechanical fracture in ($\alpha$-V$_2$O$_5$ single-crystal) positive electrode materials}
			\title{Deciphering the interplay between wetting and chemo-
					\\mechanical fracture in lithium-ion battery cathode materials}
			\maketitle
		}
		
		% Author: Please give full first and last names for authors and include * after the name of all corresponding authors
		
		{\RaggedRight
			\author{Wan-Xin Chen$\ddag$}
			\author{Luis J. Carrillo$\ddag$}
			\author{Arnab Maji}
			\author{Xiang-Long Peng}
			\author{Joseph Handy}\\
			\author{Sarbajit Banerjee*}
			\author{Bai-Xiang Xu*}
		}
		
		% Dedication
		\noindent \dedication{$\ddag$: These authors contributed equally to this work.}
		
		% Affiliations: Please provide adacemic titles (Prof. or Dr.) for all authors where applicable, and include an institutional email address for all corresponding authors
		\begin{affiliations}
			\noindent Wan-Xin Chen, Xiang-Long Peng, Bai-Xiang Xu\\
			Division Mechanics of Functional Materials, Institute of Materials Science, Technischen Universit\"at Darmstadt, Darmstadt 64287, Germany\\
			Email Address: \textcolor{blue}{xu@mfm.tu-darmstadt.de}\\
			
			\noindent Luis J. Carrillo, Arnab Maji, Joseph Handy \\
			Department of Materials Science and Engineering, Texas A$\&$M University, College Station, Texas 77843-3003, United States\\
			Department of Chemistry, Texas A$\&$M University, College Station, Texas 77842-3012, United States \\ 
			
			\noindent Sarbajit Banerjee\\ 
			Department of Materials Science and Engineering, Texas A$\&$M University, College Station, Texas 77843-3003, United States\\
			Department of Chemistry, Texas A$\&$M University, College Station, Texas 77842-3012, United States \\ 
			Laboratory for Inorganic Chemistry, Department of Chemistry and Applied Biosciences, ETH Zurich, Vladimir-Prelog-Weg 2, CH-8093 Z\"urich, Switzerland\\
			Laboratory for Battery Science, PSI Center for Energy and Environmental Sciences, Paul Scherrer Institute, Forschungsstrasse 111, CH-5232 Villigen PSI, Switzerland\\
			Email Address: \textcolor{blue}{sarbajit.banerjee@psi.ch}
		\end{affiliations}
		
	\end{spacing}
	
	%%%%%%%%%%%%%%%%%%%%
	%Abstract should be written in the present tense and impersonal style (i.e., avoid we), and be at most 200 words long
	\begin{abstract}
		
		Crack growth in lithium-ion battery electrodes is typically detrimental and undesirable. However, recent experiments suggest that stabilized fracture of cathode active materials in liquid electrolytes can increase electrochemically active surfaces, shorten diffusion pathway, enhance (de)lithiation and improve overall capacity. To decipher the fundamental couplings between electrolyte wetting and fracture evolution and evaluate their influences on macroscopic battery performance, we conducted an integrated experiment-simulation study on $\alpha$-V$_2$O$_5$ single crystals and polycrystalline NCM as model cathode materials. Despite synthesis challenges, single-crystal $\alpha$-V$_2$O$_5$ offers clearer fundamental insights than polycrystalline counterparts with grain-boundary complexities. Fracture patterns and lithiation heterogeneities on the samples were mapped using advanced \textcolor{black}{spectromicroscopy} techniques after chemical (de)lithiation cycles, exhibiting excellent agreements with simulations by the developed multiphysics model. Results reveal a mutually reinforcing interplay between wetting and fracture: (i) electrolyte infiltration at fracture surfaces enhances (de)lithiation and compositional heterogeneity; (ii) wetting influences fracture dynamics, including fracture modes, propagation distance, and directionality. The validated modelling framework is further applied to simulations on polycrystalline NCM particles under constant-current (dis)charging, highlighting the critical role of wetting in promoting fracture and improving overall capacity. This work bridges fundamental understanding of wetting–fracture coupling with practical implications for battery performance optimization via controlled fracture engineering.
		
	\end{abstract}
	
	%%%%%%%%%%%%%%%%%%%%
	% Keywords: Please provide a minimum of three and a maximum of seven keywords, separated by commas
	{
		\RaggedRight
	}
	
	% Text: Please use section headings and subheadings as specified below. For communications, all section headings apart from Experimental Section should be removed
	% Please make the first reference to a display item bold: \textbf{Figure 1}
	% Do not abbreviate Figure, Equation, etc.; display items are always singular, i.e., Figure 1 and 2.
	% Equations are always singular, i.e., Equation 1 and 2, and should be inserted using the {equation} environment, not as graphics
	% Please do not use footnotes in the text, additional information can be added to the Reference list.
	
	\section{Introduction}
	
	\hspace{4mm}
	Lithium-ion batteries (LIBs) have become the dominant energy storage technology for electronics, electromobility, and emerging grid-scale applications.\cite{Whittingham2020,Cano2018} However, their widespread adoption continues to be hindered by chemo-mechanical fracture induced performance degradation and capacity fade,\cite{Lewis2019,Santos2022,Xing2022} which limits their long-term stability and reliability. Understanding mechanisms of fracture within the battery and their intricate couplings with chemo-mechanical behavior is critical for developing next-generation LIBs with improved performance and longevity. In this paper, we focus on the cathode materials of LIBs, which are typically oxide ceramics that oftentimes exhibit compositional heterogeneities during Li-ion insertion and extraction, arising from concentration gradients, phase transformations and the anisotropy of ion diffusion. \cite{Augustyn2020ACSEnergyLetter,Santos2022_2,mistry2022asphericity} These processes are accompanied by lattice expansion and contraction, leading to the development of mechanical stress and significant crack formation over repeated (de)lithiation cycles, e.g., interface delamination between the electrode and solid electrolyte,\cite{delamination,Liu2025NanoEnergy} inter-granular (interface) damage, and trans-(intra-)granular (bulk) fracture within particles of the active material.\cite{Yan2017NC,Trevisanello2021AEM,AFM1_crack,AFM2_crack}  Crack formation and material damage ultimately lead to electrode fragmentation, electrical isolation of active material, loss of ionic percolation pathways, and overall global capacity fade and performance degradation of the LIBs in the long-term cycling.\cite{Wang2019,2020review,AFM3_design} As a result, traditional battery design strategies have largely focused on suppressing or mitigating cracks formation.
	
	However, recent studies have suggested that fracture processes in lithium host materials can, under specific conditions, be leveraged to enhance lithium transport, improve charge-transfer kinetics, and ultimately contribute to better electrochemical performance. In a typical polycrystalline secondary cathode particle consisting of primary grains and grain boundaries,\cite{Trevisanello2021AEM,Furat2021JPS} such as LiNi$_x$Mn$_y$Co$_{1-x-y}$O$_2$ (NMC or NCM), lithiation and delithiation are conventionally understood to initiate at the surfaces of sintered secondary aggregates, which are in contact with the liquid electrolyte, and subsequently propagate into the interiors through a combination of bulk and grain boundary diffusion.\cite{Tsai2018EES,Xu2019JMPS} Recently, work by Janek, Li, Zhao, etc., \cite{Ruess2020JES,Trevisanello2021AEM,Li2023EES,Han2024JPS,Kejie2018NanoEnergy,Peter2022AEM,Deng2020Wetting} indicate that in liquid-electrolyte-based cells, incipient inter-granular fracture surfaces at grain boundaries can introduce an additional electrochemical aspect within the electrode, which further enhances the transport of lithium within the active particles and improves charging capacity by increasing the active surface area and shortening diffusion pathways as compared to the initial undamaged regime.\cite{Li2023EES,Han2024JPS} These results challenge conventional assumptions about fracture in battery cathode materials, suggesting that under controlled conditions, crack formation and stabilization may be beneficial. Despite growing interest, the coupled mechanisms of electrolyte wetting and fracture evolution remain poorly understood. Experimental studies have yet to elucidate the direct impact of electrolyte infiltration on crack nucleation and propagation dynamics. Furthermore, most investigations have focused on macroscopic battery performance metrics (for example, overall specific capacity) rather than resolving local compositional heterogeneities induced by wetting. This challenge is further compounded in polycrystalline electrode particles, where the microstructural complexities introduced by grain boundaries hinder direct tracking of diffusion heterogeneities and crack evolution.\cite{Trevisanello2021AEM,Deng2023SinglePoly,Homlamai2022CC} 
	As a result, the absence of direct experimental mapping of electrolyte infiltration and its influence on fracture propagation presents a critical gap, hindering the ability to accurately uncover the fundamental chemo-mechanically coupled behaviors in lithium-ion battery electrodes.
	
	In contrast, although synthesizing single-crystal electrode materials remains challenging, their well-defined crystallographic structures provide a powerful platform for probing the fundamental principles of chemo-mechanical coupling.\cite{Langdon2021,Zeng2022,Santos2022_2,Lu2022Small} This distinct advantage makes them an appealing alternative to conventional polycrystalline counterparts with grain boundary complexities.
	%the influence of crystallographic-structure-dependent chemo-mechanical properties
	In this study, we systematically investigate the complex interplay between electrolyte wetting and fracture evolution in cathode materials, and evaluate their impact on macroscopic battery performance, using both single-crystal and polycrystalline cathode systems as model platforms.
	To elucidate the fundamental coupling mechanisms, we employ $\alpha$-V$_2$O$_5$ single crystals with minimal microstructural complexity interference. An integrated experimental and simulation approach is adopted: multimodal electron and X-ray microscopy directly map cracking patterns and electrolyte-wetting-induced lithium heterogeneity following controlled chemical lithiation–delithiation cycles, in parallel, we develop a comprehensive multi-physics simulation framework that uniquely captures (i) the concurrent evolution of interfacial and bulk fractures within a unified model,\cite{Chen2024JMPS} and (ii) the coupled effects of electrolyte infiltration along both fracture types. The proposed model addresses limitations in earlier models, which typically accounted for only one type of fracture-induced infiltration.\cite{Zhao2016CMAME,Xu2016GAMM,Emilio2023CMAME,Miehe2016IJNME,Han2024JPS}
	Building on the established wetting–fracture coupling mechanisms revealed in single-crystal cathode materials, the simulation framework is subsequently extended to polycrystalline cathode particles that have been experimentally studied in prior work.\cite{Ruess2020JES,Trevisanello2021AEM} In particular, we numerically investigate the role of electrolyte wetting and its coupling with fracture propagation in polycrystalline NCM particles under electrochemical (dis)charging conditions. Our analysis highlights how these multiphysical couplings affect macroscopic battery responses, including voltage behavior and capacity performance, thus bridging fundamental understanding with practical device implications.

	\section{Results and discussions}
	
	\subsection{Experimental and simulation methods}
	
	\begin{figure}%[htbp]
	\centering
	\includegraphics[width=1.0\textwidth]{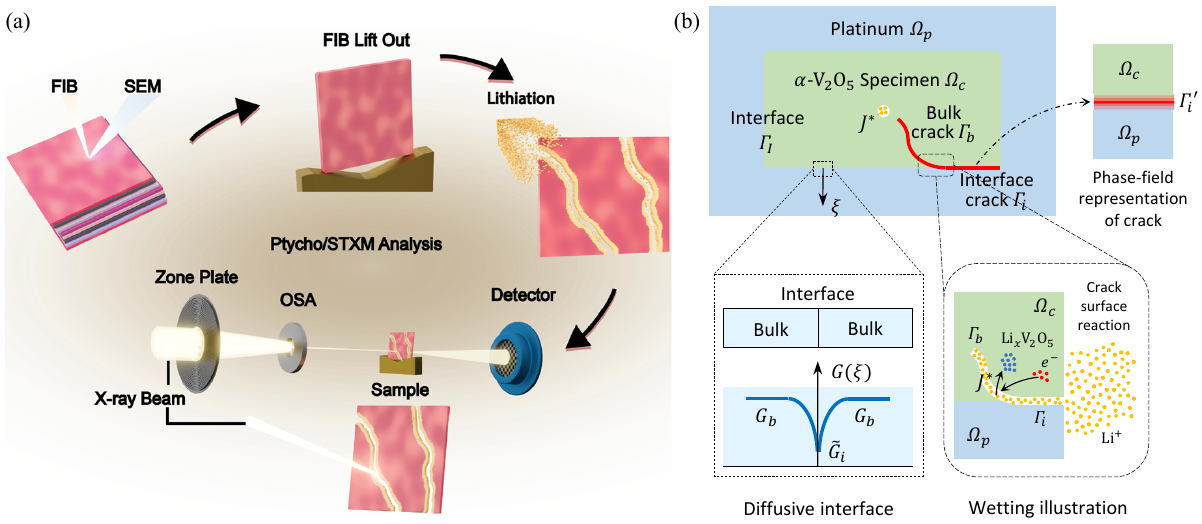}
	\caption{Schematic illustrations of experimental (a) and simulation (b) configurations. Panel (a) illustrates the experimental workflow, starting from a single crystal of $\alpha$-V$_2$O$_5$ prepared for etching using FIB and SEM, 
	%\textcolor{red}{including the creation of a circular notch/hole in the center within the sample. 
	%The lamella sample is further assembled with a Pt frame} 
	which is subjected to (de)lithiation cycles. Next, the damage patterns are imaged using SEM/TEM. Concurrently, the lithium composition is mapped using scanning transmission X-ray microscopy (STXM).\cite{Santos2022} Panel (b) highlights key aspects of the simulation method. The computational domain comprises the $\alpha$-V$_2$O$_5$ lamella $\varOmega_c$ and platinum frame $\varOmega_p$; the phase-field fracture model is adopted to model (de)lithiation induced bulk and interface cracking evolutions. Two novel points are emphasized in two zoomed call-outs: (i) diffusive treatment of the fracture energy to simulate interface delamination,\cite{Chen2024JMPS} (ii) electrochemical reaction and chemical flux at bulk and interface cracking surfaces arising from wetting effects.\cite{Zhao2016CMAME, Xu2016GAMM,Emilio2023CMAME}
	}
	\label{fig:configuration}  
	\end{figure}
	
	\hspace{4mm}
	Figure~\ref{fig:configuration} (a) briefly depicts an overview of the experimental workflow on single-crystal cathode materials. In this study, $\delta$-Li$_{0.7}$V$_2$O$_5$ powders were synthesized via a solvothermal process,\cite{handy2022li} followed by annealing and controlled crystallization to produce macroscopic sub-mm-sized single crystals. These were subsequently topochemically deintercalated to obtain layered $\alpha$-V$_2$O$_5$ single crystals.\cite{Santos2022,Santos2022_2} A focused ion beam (FIB) was then used to etch the millimeter-sized single crystal layers, forming micrometer-scale lamella. %(Figure~\ref{fig:old-sample} (b), (c) \textcolor{red}{I am confused by this. Is this supposed to say Figure 1? If so, the "a,b,c" labels aren't present in the figure} ). 
    The prepared lamella was subjected to chemical lithiation and delithiation cycles using \textit{n}-butyllithium and NOBF$_4$, respectively. During this process, complex chemo-mechanical interactions led to fracture propagation, which \textcolor{black}{were} characterized using scanning \textcolor{black}{electron microscopy} and transmission electron microscopy (SEM/TEM). Finally, lithium distribution was mapped through scanning transmission X-ray microscopy (STXM) and ptychography.\cite{Santos2022} Further details on the experimental procedures are provided in the Supporting Information (Section 1).
	
	To understand the wetting-fracture coupling mechanism and their influences on multi-physics behaviours in the experiments, we developed the model featuring following key aspects: 
	%(i) displacement $\boldsymbol{u} (\boldsymbol{x}, t)$ and the corresponding strain $\bfepsilon = \bfepsilon_{e} + \bfepsilon_{c}$ are adopted to model the mechanical process, where 
	%the chemical strain $\boldsymbol{\epsilon}_{c}$ is considered in the mechanical process to model the concentration-variation-induced deformation and stress; 
	(i) cohesive phase-field fracture model is adopted to simulate lithium intercalation/extraction induced fracture evolution.\cite{Chen2024JMPS,Wu2017JMPS} In particular, the unified framework enables the concurrent simulation of bulk and interfacial fractures by appropriately assigning fracture energy properties within the domain, for instance, a diffusive interface with an effective interface fracture energy $\tilde{G}_i$ is depicted in Figure~\ref{fig:configuration} (b).\cite{Chen2024JMPS} (ii) lithium concentration $c(\boldsymbol{x}, t)$ satisfies the mass conservation law $\dot{c} + \nabla \cdot \boldsymbol{J} = Q$, with source term $Q$ introduced at the (diffuse) fracture surfaces to model wetting phenomenon, e.g., $Q = 2G(\boldsymbol{x}) / G_i \gamma(d, \nabla d) J^{\ast}$ and $Q = 2G(\boldsymbol{x}) / G_b \gamma(d, \nabla d) J^{\ast} = 2 \gamma(d, \nabla d) J^{\ast}$ are properly defined respectively for interface and bulk fracture induced wetting, which extends electrochemical reaction and corresponding chemical flux $J^{*}$ to fracture surfaces beyond sample's external boundary.
	% simulation capabilities to interface-delamination-induced wetting, previously limited to bulk fracture in classical phase-field models \cite{Zhao2016CMAME,Miehe2016IJNME,Emilio2023CMAME}. 
	%
	It is worth noting that the aforementioned methodology, while demonstrated for $\alpha$-V$_2$O$_5$ single crystals, is also applicable to subsequent simulations of polycrystalline cathode materials, where both trans-granular (bulk) cracks within grains and inter-granular (interface) fracture along grain boundaries can concurrently happen.\cite{Chen2024JPS} A comprehensive description of the thermodynamically consistent model derivation, simulation setup, adopted parameters, and other relevant aspects of the modeling framework is provided in the Supporting Information (Section 2 and 3).
	
	\subsection{Results on single-crystal $\alpha$-V$_2$O$_5$ cathode materials}
	
	\hspace{4mm}
	This section explores the fundamentals of chemo-mechanical coupling between wetting phenomenon and fracture evolution in cathode materials through integrated experimental and simulation approaches. Single-crystal $\alpha$-V$_2$O$_5$ samples were selected to minimize microstructural complexity and isolate the underlying coupling mechanisms. Following controlled chemical lithiation-delithiation cycles, lithium concentration profiles and fracture patterns were characterized via electron and X-ray microscopy. Corresponding simulations reproduced the experimental conditions and yielded results in strong agreement, validating the modeled coupling behaviours.
	
	\subsubsection{Experimental $\&$ simulation investigations on Sample-1}
	
	\begin{figure}[htbp]
		\centering
		\includegraphics[width=0.825\textwidth]{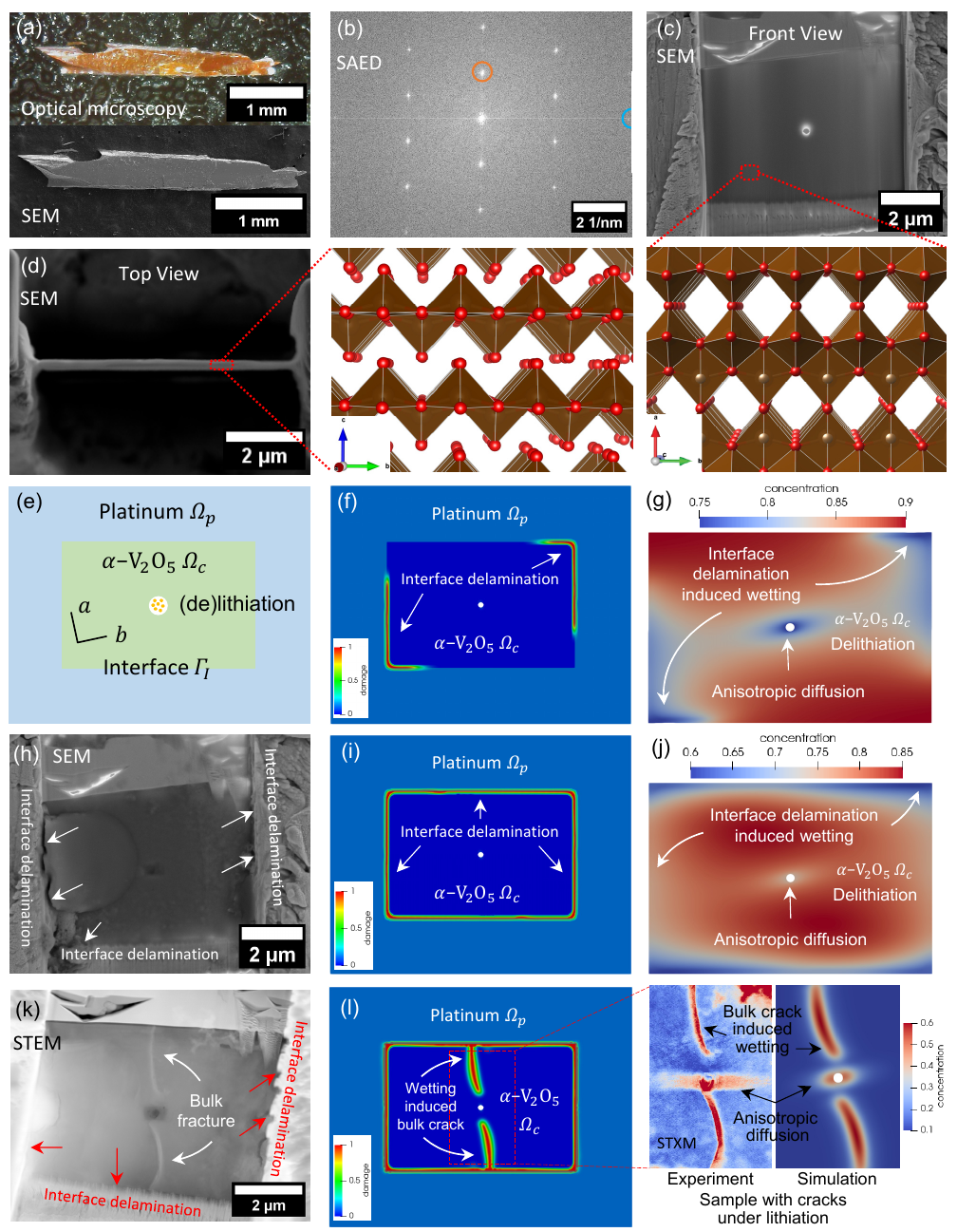}
		\caption{Comparison of experiment and simulation results of the Sample-1. 
			Panel (a) shows optical microscopy and SEM images of the single crystal used for FIB lift out. 
			Panel (b) illustrates the SAED pattern of the lamella ([200] in orange and [030] in blue). 
			Panels (c) and (d) show the $\alpha$-V$_2$O$_5$ lamella and platinum frame from the front and top view, respectively; the callouts show the crystallographic representation of the $\alpha$-V$_2$O$_5$ lamella along the corresponding perspectives. 
			Panel (e) illustrates the computational domain of the simulation, i.e., including platinum $\varOmega_p$ and $\alpha$-V$_2$O$_5$ $\varOmega_c$ with in-between interface. 
			Panels (f) and (g) show the resulting cracks and concentration map upon delithiation. 
			Panels (h) and (i) compare the fracture results at the end of one (de)lithiation cycle, they are respectively obtained from experiment (SEM image) and simulation; 
			Panel (j) shows the corresponding concentration map. 
			Panels (k) and (l) compare the ultimate failure mode of the sample obtained respectively from experiment (STEM micrograph) and simulation. 
			Panel (l) also provides a comparison of concentration map upon lithiation based on \textcolor{black}{X-ray ptychography} imaging and simulation methods in the call-out. 
			%\textbf{A.} Image of single crystal used for FIB lift out under optical micscroscopy and SEM; \textbf{B.} crystallographic representation of the front side of the FIB lamella; \textbf{C.} crystallographic representation of the top side of the FIB lamella; \textbf{D.} SAED of the FIB lamella taken in the front of the lamella and showing the indexed [200] in orange and [030] in blue; \textbf{E.} shows the SEM micrograph of the front side of the FIB lamella before lithiation; \textbf{F.} shows the SEM micrograph of the top side of the FIB lamella before lithiation; \textbf{G.} shows the SEM micrograph of the front side of the FIB lamella after lithiation where the delamination is highlighted with the number 1; \textbf{H.} shows the STEM micrograph of the front tide of the FIB lamella after lithiation where the delamination is highlighted with the number 1 and the internal cracks are highlited with the number 2; Subfigure (9) illustrates the simulation domain, including the platinum $\varOmega_p$, the Li$_x$V$_2$O$_5$ active cathode material $\varOmega_c$ and the in-between interface $\varGamma_I$, as well as the orientation of the $\varOmega_c$. Subfigure (10-12) shows the evolution of cracks in the structure, at the beginning the interface delamination (denoted as mode-1) propagates, with the interface bulk crack (denoted as mode-2) growing subsequently. Subfigure (13-14) show the concentration at two moments. Subfigure (15) compare the wetting induced local high concentration results by simulation (left) and experiment (right) methods.
		}
		\label{fig:old-sample}  
	\end{figure}
	
	\hspace{4mm}
	Figure~\ref{fig:old-sample} presents the experimental and simulation setups, along with the corresponding results for Sample-1. A millimeter-sized $\alpha$-V$_2$O$_5$ single crystal was grown, as shown in the optical microscopy and SEM images displayed in Figure~\ref{fig:old-sample} (a), whose single-crystalline character and single domain nature is confirmed by selected area electron diffraction (SAED) result (Figure~\ref{fig:old-sample} (b)). The FIB etched micrometer-sized Sample-1 is displayed in Figure~\ref{fig:old-sample} (c) from the front view and in Figure~\ref{fig:old-sample} (d) from the top view, with crystal structures shown in the corresponding zoomed call-outs. As shown in Figure~\ref{fig:old-sample} (c), a circular notch (hole) is generated using FIB at the sample's center to introduce an initial diffusion channel, and the sectioned lamella is positioned within a Pt frame deposited using FIB-SEM. The surrounding Pt effectively blocks (de)intercalation from the in-plane four edges, meanwhile, (de)lithiation at out-of-plane directions is also impeded due to extremely low diffusivity at $c$-axis direction. Moreover, mechanical interactions between the two materials are facilitated along their interface. Consistent with experiment configurations, the 2D computational domain in Figure~\ref{fig:old-sample} (e) is considered.  A prescribed chemical flux $J^{\ast}$ is applied on the internal hole surface to mimic lithiation condition in the experiment ($-J^{\ast}$ for delithiation). In the simulation, the $b$-axis with higher diffusivity is assumed to be oriented with deviation by $10^{\circ}$ from the horizontal axis, which is supposed to remain a valid approximation, considering potential misalignment or inclination in the actual geometry of the assembled $\alpha$-V$_2$O$_5$ lamella and Pt frame.
	
	Upon (de)lithiation, varying lithium concentration induces volume changes in the host material $\alpha$-V$_2$O$_5$, leading to the development of mechanical stress, especially, tensile stresses manifest during delithiation,\cite{Chen2024JPS,Klinsmann2015JES} which induce fracture initiation within the structures. Figure~\ref{fig:old-sample} (f) presents the simulated interface delamination arising from the mechanical mismatch. Specifically, the $\alpha$-V$_2$O$_5$ lamella undergoes lattice shrinkage due to delithiation, whereas the platinum frame exhibits negligible volume change. As such, delamination initiates at the left and right edges due to the preferred lithium diffusion pathway along $b$-axis (see diffusion in Figure~\ref{fig:old-sample} (g)).
	%Furthermore, due to a preference in lithium diffusion along the $b$-axis, which can be confirmed in the simulated concentration map in \cref{fig:old-sample} (f) as well, delamination initiates at the left and right segments of the interfaces.
	Concurrently, wetting phenomenon is initiated, wherein the liquid electrolyte infiltrates the newly generated fracture surfaces, activating (de)lithiation reactions at the exposed areas. As a result, lower lithium concentration near the delamination zone during delithiation is observed in Figure~\ref{fig:old-sample} (g). Subsequently, the interface delamination progresses toward the top and bottom interfaces, as observed in both the experimental SEM image (Figure~\ref{fig:old-sample} (h)) and the simulation result (Figure~\ref{fig:old-sample} (i)), which exhibit a high degree of agreement. Figure~\ref{fig:old-sample} (j) presents the concentration map after one cycle obtained by the simulation. Indeed, wetting plays a pivotal role in the global concentration heterogeneity, i.e., with the completed delamination and the activated wetting, four edges of the lamella present lower lithium concentration, especially the top and bottom edges due to lower diffusivity along $a$-axis, where high tensile stress is developed to further drive the propagation of internal bulk cracks, as seen in the STEM image (Figure~\ref{fig:old-sample} (k)) and simulation result (Figure~\ref{fig:old-sample} (l)), which exhibit excellent agreement regarding the failure mode. The call-out in Figure~\ref{fig:old-sample} (l) presents a comparison of lithium concentration maps of the sample under lithiation, demonstrating excellent consistency with the STXM-derived concentration map \cite{Santos2022,Luo2022PNAS} (see details in Section 1.5 in the Supporting Information) and simulation result. Both reveal electrolyte wetting at the internal bulk cracks, leading to locally elevated lithium concentrations during lithiation and highlighting its interplay with fracture evolution process; besides, locally high concentration at $b$-axis direction near the central hole due to anisotropic diffusion, can also be observed.
	
	\subsubsection{Simulation investigations comparing with/without wetting}
	
	\begin{figure}[htbp]
		\centering
		\includegraphics[width=1.0\textwidth]{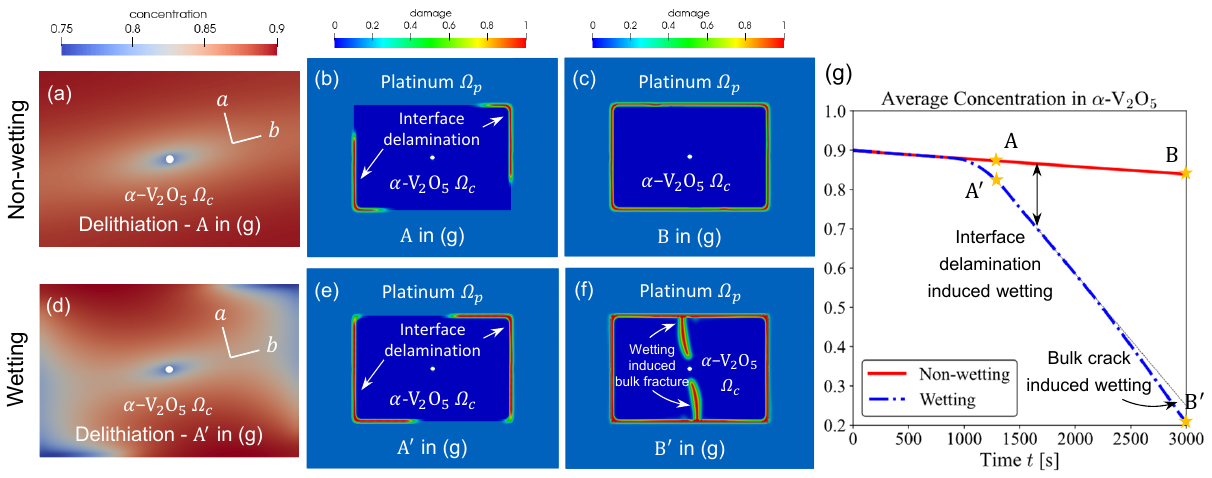}
		\caption{Comparison of simulation results between wetting and non-wetting regimes under delithiation conditions. 
		Panels (a) and (d) present the concentration profiles, 
		Panels (b), (c), (e) and (f) illustrate the fracture evolution processes, with interface delamination observed in both wetting and non-wetting regimes, while bulk cracks exist only in the wetting regime. 
		Panel (g) depicts the temporal evolution of the average lithium concentration within $\alpha$-V$_2$O$_5$, comparing cases with and without the wetting effect. %\textcolor{red}{Can you move the "(g)" to the top left of the graph? it would make it easier to see}
		}
		\label{fig:comparison-with-without-wetting}  
	\end{figure}
	
	%\hspace{4mm}
	To further elucidate the coupling between electrolyte wetting and chemo-mechanical fracture, comparative simulations are conducted with/without considering the wetting phenomenon, serving as a computational complement to the experimental observations.
	%(ii)
	%These simulation results complement the experimental findings, revealing the intricate chemo-mechanical coupling mechanisms and the reciprocal reinforcement between fracture evolution and diffusion dynamics.
	Figure~\ref{fig:comparison-with-without-wetting} compares the concentration results and fracture evolution processes for wetting and non-wetting regimes under delithiation conditions, other setups of simulations are similar to the previous example. With the continuous deintercalation of lithium, mechanical mismatch between the $\alpha$-V$_2$O$_5$ lamella and the platinum frame leads to interface delamination in both wetting and non-wetting regimes (see Figure~\ref{fig:comparison-with-without-wetting} (b) and (e)). In the wetting regime, as shown in Figure~\ref{fig:comparison-with-without-wetting} (d) when comparing with Figure~\ref{fig:comparison-with-without-wetting} (a), electrolyte infiltration at freshly formed interface cracks lowers the lithium concentration near delaminated regions; a higher decreasing rate of average lithium concentration is observed in Figure~\ref{fig:comparison-with-without-wetting} (g). As a fully coupled system, enhanced delithiation accelerates fracture propagation, as evidenced by larger delamination distances in the wetting regime (Figure~\ref{fig:comparison-with-without-wetting} (e)) compared to the non-wetting case (Figure~\ref{fig:comparison-with-without-wetting} (b)). Figure~\ref{fig:comparison-with-without-wetting} (c) and (f) compare distinct ultimate fracture modes with and without considering wetting. In the wetting regime, chemical flux and enhanced delithiation at delaminated surfaces induce tensile stress within $\alpha$-V$_2$O$_5$ lamella, which drives further propagation of internal bulk cracks. In contrast, the absence of wetting prevents the propagation of bulk cracks, which underscores its crucial role in fracture evolution. Once initiated, bulk cracks further enhance delithiation by exposing new surfaces and pathways, thereby accelerating concentration changes and steepening the decline in average lithium concentration, as can be seen in Figure~\ref{fig:comparison-with-without-wetting} (g), where different slopes could be observed between dash-dot blue line (with bulk crack and interface delamination induced wetting) and black reference line (with interface delamination induced wetting).
	
	\subsubsection{Simulation investigations on the influence of crystal orientation}
	
	\begin{figure}[htbp]
	\centering
	\includegraphics[width=0.9\textwidth]{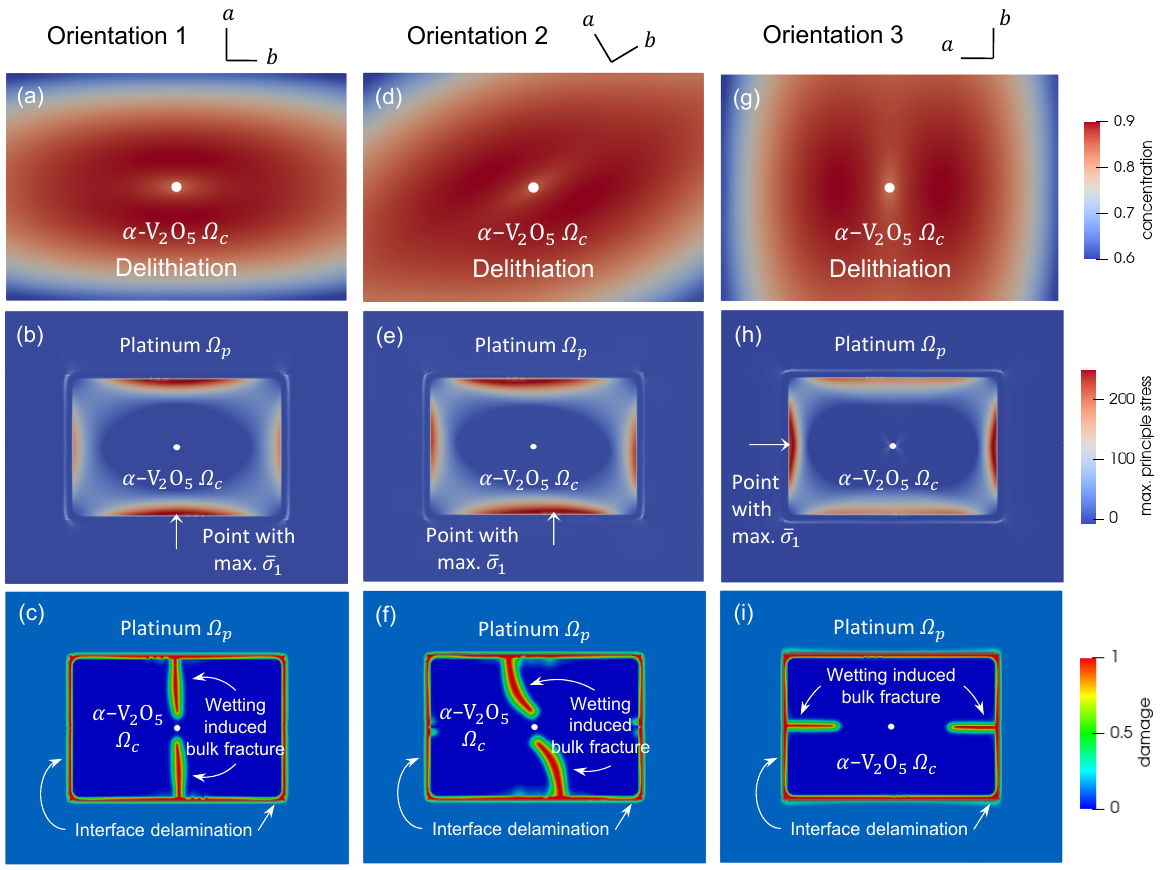}
	\caption{Simulations investigating the influence of Li-ion diffusion pathway orientation at the delithiation step. Panels (a), (d) and (g) show Li concentration maps of the active materials across different orientations,  following complete interface delamination under delithiation conditions. Panels (b), (e) and (h) display the stress distribution of active materials and Pt under various orientations. Panels (c), (f) and (i) illustrate ultimate fracture patterns in the structure under different orientations.
	}
	\label{fig:study-orientation}  
	\end{figure}
	
	\hspace{4mm}
	While wetting plays a critical role in the fracture evolution of cathode active materials, lithium-ion diffusion anisotropy—governed by the underlying crystal structure—can further couple with wetting effects, significantly influencing stress development and the directional formation of internal bulk fractures.\cite{Han2024JPS,Parija2016CM} In this study, we employ numerical simulations to systematically investigate these coupled effects and provide complementary insights to the experimental findings. Figure~\ref{fig:study-orientation} presents concentration maps, stress distributions, and crack phase-field results for three different crystal orientations, incorporating anisotropic diffusivity.\cite{Ma2013JPD,Xu2021ACS} Notably, in-plane elasticity along the $a$- and $b$-axes shows minimal variation from the out-of-plane ($c$-axis) direction, whose anisotropic effects are not considered here.\cite{Reed2020PRB,Jachmann2005SSC} Figure~\ref{fig:study-orientation} (a), (d), and (g) compare concentration profiles under delithiation condition after the completed interface delamination. A narrow strip from the central notch along the preferred $b$-axis with lower concentration forms due to lithium deintercalation. More interestingly, wetting has a greater impact, depleting lithium along all four edges of the $\alpha$-V$_2$O$_5$ sample. Edge segments at the $a$-axis-direction show pronounced depletion (or enrichment under lithiation) due to limited diffusion, where the highest tensile stress driving the formation of internal bulk crack is concentrated, as shown in Figure~\ref{fig:study-orientation} (b), (e), and (h). Figure~\ref{fig:study-orientation} (c), (f), and (i) illustrate the fracture patterns, showing that internal bulk cracks predominantly form perpendicular to the preferred diffusion pathway ($b$-axis), resulting from the direction of generated maximum principal stress. These results underscore the significant role of crystal structure and diffusion pathway orientation in governing the directionality of crack evolution within the lamella.
	
	\subsubsection{Experimental $\&$ simulation investigations on Sample-2}
	
	\begin{figure}[htbp]
	\centering
	\includegraphics[width=0.9\textwidth]{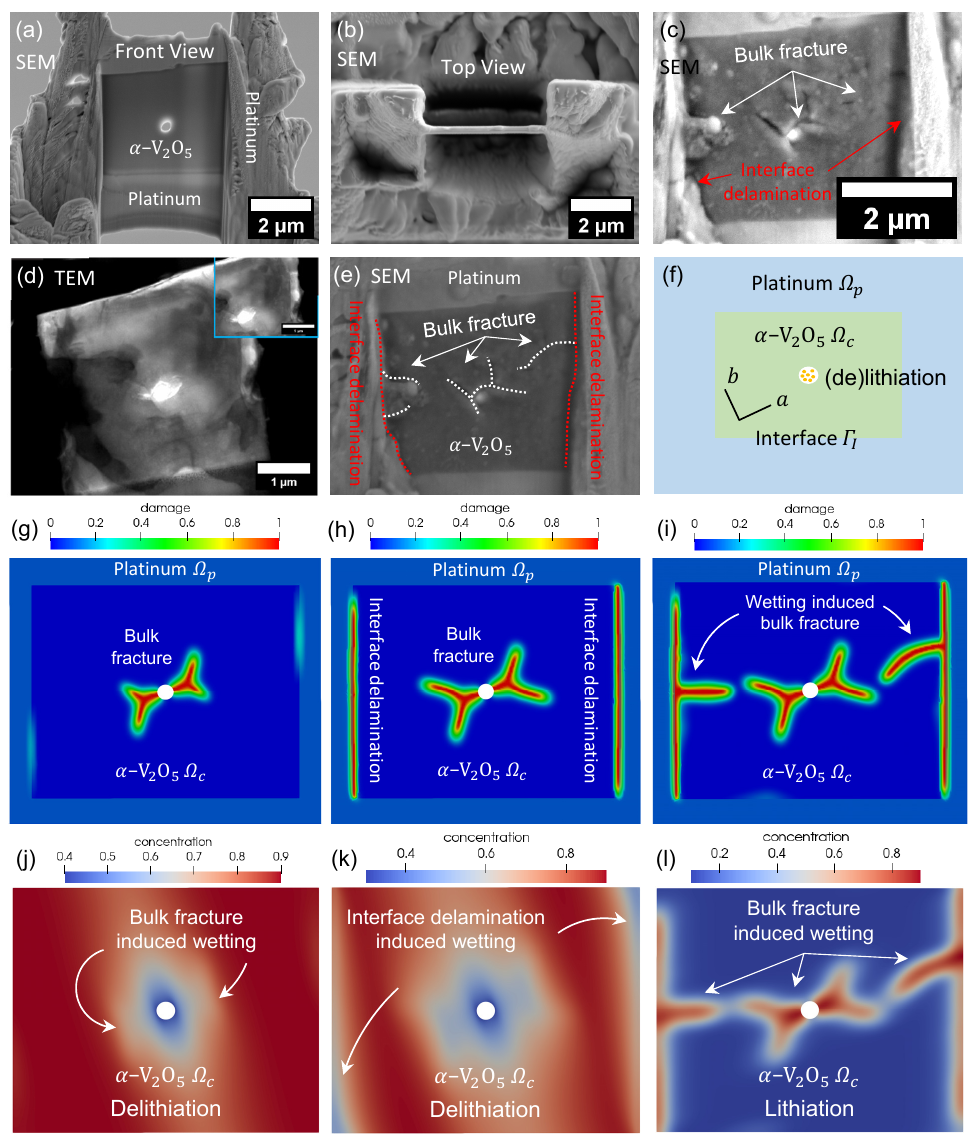}
	\caption{Comparison of experiment data and simulation results of the Sample-2 with extended chemical (de)lithiation cycles. Panels (a) and (b) show the $\alpha$-V$_2$O$_5$ lamella and platinum frame from the front and top view, respectively. Panels (c) and (d) show the SEM and TEM images of the sample after (de)lithiation cycles, respectively; Panel (e) displays the SEM image of the polished sample after 6 lithiation/delithiation cycles. Panel (f) illustrates the computational domain in the simulation, i.e., including platinum $\varOmega_p$ and $\alpha$-V$_2$O$_5$ $\varOmega_c$ with an in-between interface. Panels (g), (h) and (i) show the evolution of fracture in the sample, and Panels (j), (k) and (l) illustrate the corresponding concentration maps.%\textcolor{red}{Is there a reason the panel headers are not alphabetized?}
	}
	\label{fig:new-sample}  
	\end{figure}
	
	\hspace{4mm}
	Building on the identified reciprocal reinforcement mechanism between fracture evolution and wetting-enhanced diffusion dynamics, we next analyze the experimental and simulation results for Sample-2, where extended cycling leads to the formation of more severe cracks. A new sectioned single-crystal lamellae, with FIB-generated center hole and surrounding Pt frame, is shown in Figure~\ref{fig:new-sample} (a) and (b) from front and top perspectives, respectively.
	The sample underwent six lithiation and delithiation cycles. Repeated cycling led to an increased formation of byproduct Li$_2$O (see details in Section 1.3 in the Supporting Information),\cite{canepa2017odyssey,Wang2016PRB} which appears as debris in the SEM image (Figure~\ref{fig:new-sample} (c)) and the TEM image (Figure~\ref{fig:new-sample} (d)). The SEM image of polished sample with cracks formation during cycles is shown in Figure~\ref{fig:new-sample} (e).
	The simulation follows the approach used for Sample-1, with differences including the orientation shown in Figure~\ref{fig:new-sample} (f) and a larger chemical flux ($2J^{\ast}$) applied to the internal hole surface. The resulting fracture patterns and concentration profiles over time are sequentially presented in Figure~\ref{fig:new-sample} (g)–(l).
	
	The increased delithiation flux at the hole surface, combined with the mechanical constraint from the Pt frame, induces tensile stress near the notch, promoting bulk crack nucleation and evolution in Figure~\ref{fig:new-sample} (g). Simultaneously, electrolyte infiltration into the bulk crack surface results in a decrease in lithium concentration near the hole, as seen in Figure~\ref{fig:new-sample} (j). As delithiation progresses, evolution of bulk cracks continues with distinctive branching patterns observed in Figure~\ref{fig:new-sample} (h), which are in excellent agreement with the experimental results in Figure~\ref{fig:new-sample} (c); meanwhile, delamination occurs at the interface between the $\alpha$-V$_2$O$_5$ lamella and the Pt frame, as shown in Figure~\ref{fig:new-sample} (h) and confirmed by the TEM image in Figure~\ref{fig:new-sample} (d). Concurrently, lithium depletion is increased in the delaminated region, causing a more rapid decline of lithium concentration, as observed in Figure~\ref{fig:new-sample} (k). Over multiple cycles, new bulk cracks initiate at the left and right edges of the lamella, propagating inward due to the increasing wetting flux at the delaminated surface, see comparison between experimental (Figure~\ref{fig:new-sample} (e)) and simulated (Figure~\ref{fig:new-sample} (i) crack patterns. As a result, newly formed crack surfaces within the $\alpha$-V$_2$O$_5$ become wetted, leading to locally elevated lithium concentrations near bulk fracture surfaces under lithiation (Figure~\ref{fig:new-sample} (l)). The coupled effects of wetting and fracture evolution are systematically explored through integrated experiments and simulations, using single-crystal cathode materials to isolate the role of wetting without the interference from microstructural complexity. The results consistently capture key failure modes—including interfacial delamination, delithiation-induced branched bulk cracks, and wetting-driven bulk fracture—highlighting the critical role of wetting in fracture behaviours of cathode materials.
	
	\subsection{Results on ploycrystalline NCM cathode materials}
	
	\hspace{4mm}
	Building upon the established mutually reinforcing coupling between fracture propagation and wetting phenomenon from single-crystal cathode materials, this section extends the modeling framework to electrochemical charging regimes within the polycrystalline active materials. In this study, we specifically investigate the critical role of wetting phenomena in governing multiphysical responses—such as chemical diffusion, fracture evolution, voltage characteristics, and capacity performance—and validate the applicability of the wetting–fracture coupling mechanism in polycrystalline material systems through comparison with previously reported experimental results. The electrochemical charging process is modeled using Butler–Volmer kinetics under charge conservation, which govern the local current density and lithium flux at the surfaces, and facilitate the transport of lithium ions from the electrolyte into the intercalated state within the cathode material. In this study, a constant current (galvanostatic) charging condition is imposed, whereby the total lithium-ion flux across all active surfaces — including the external particle boundary and the newly formed crack surfaces — remains constant throughout the charging/discharging cycle. For detailed mathematical formulations, the reader is referred to the Supporting Information (Section 2.1).
	
	\subsubsection{Electro-chemo-mechanical-fracture simulation benchmark example}
	
\begin{figure}[htbp]
	\centering
	\includegraphics[width=0.9\textwidth]{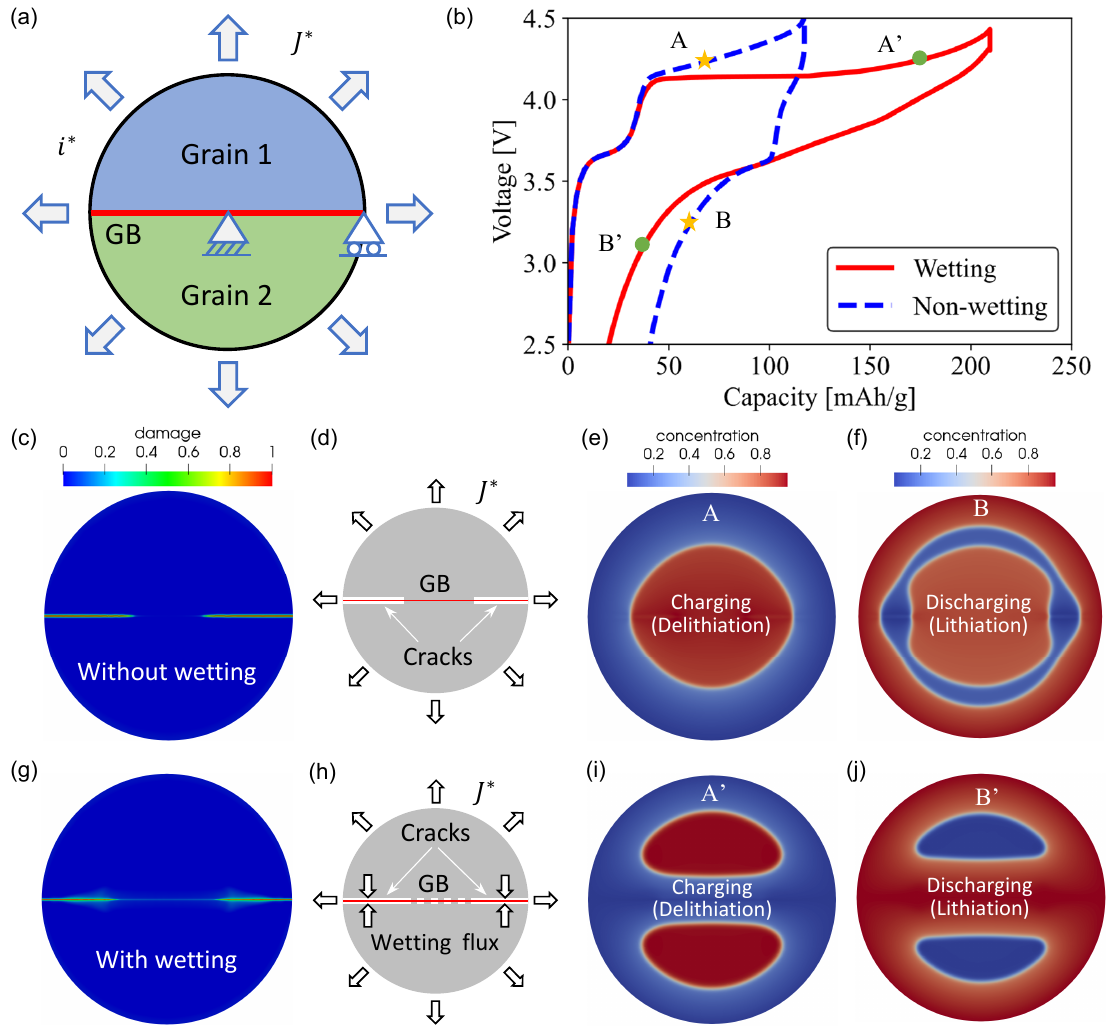}
	\caption{
		Benchmark simulation comparing electro-chemo-mechanical-fracture behaviours under wetting and non-wetting regimes. Panel (a) illustrates the simulated geometry comprising two grains separated by a grain boundary, properly defined boundary conditions for multi-physical processes are also illustrated. Panel (b) compares voltage-capacity profiles during the first constant-current charge–discharge cycle for both regimes. Panels (c) and (g) show the simulated particle fracture patterns without/with wetting effect, respectively. Panels (d) and (h) illustrate the active surfaces for electrochemical reactions and lithium ingress in the presence of fractures, without/with wetting effect, respectively. Panels (e) and (f) show the lithium concentration maps in the charging and discharging stages in the non-wetting regime, extracted at time points marked in Panel (b). Panels (i) and (j) show the lithium concentration maps in the charging and discharging stages in the wetting regime, corresponding to the time points marked in Panel (b). %\textcolor{red}{Is there any significance to the white lines appearing in c, e, f, i, and j?}
	}
	\label{fig:benchmark}  
	\end{figure}

	\hspace{4mm}
	As shown in Figure~\ref{fig:benchmark}(a), the modeled benchmark geometry consists of upper and lower grains separated by a grain boundary (GB) with relatively lower fracture energy, which serves as a preferential path for crack propagation. Butler–Volmer kinetics are applied at the external surface to simulate a single charge–discharge cycle under a constant 1C current, starting from an initial lithium concentration of $\tilde{c}_0 = 0.99$. The discharging stage begins once the upper cut-off voltage of 4.5 V is reached during charging and continues until the particle reaches the lower cut-off voltage of 2.5 V. Mechanical boundary conditions include fixed displacements in both horizontal and vertical directions at the central point, with vertical displacement constrained at the right ending point.\cite{Chen2024JPS} To facilitate benchmarking of the coupled electro-chemo-mechanical-fracture model and to isolate fundamental effects of wetting phenomena, the material is assumed to exhibit isotropic properties, such as diffusivity and elastic stiffness.
	
	Figure~\ref{fig:benchmark} (c) presents the voltage–capacity curves for a constant-current charge–discharge cycle, comparing scenarios with and without the wetting phenomenon. At the onset of the charging (delithiation) stage, lithium ions continuously exit the particle through the external surface, resulting in a decrease of lithium concentration and a corresponding increase of voltage. During this initial phase, the curves for both cases overlap due to the absence of crack formation. Subsequently, delithiation-induced tensile stress drives crack propagation from the external surface into the interior, preferentially along the grain boundary (GB) due to its lower fracture energy.\cite{Chen2024JMPS} This crack evolution activates additional electrochemical reactions and lithium flux along the newly formed internal surfaces, which in turn promotes further crack propagation within the particle interior, as shown in the fracture patterns in Figures~\ref{fig:benchmark}(g), whereas crack growth is impeded in the interior under compressive stress in the absence of wetting (see Figures~\ref{fig:benchmark}(c)). Notably, Figure~\ref{fig:benchmark}(b) reveals that the presence of wetting significantly alters the electrochemical behavior; specifically, a higher overall specific capacity is achieved under the same upper cut-off voltage in the charging process when wetting is considered. This result is in strong agreement with experimental observations reported by Ruess et al.,\cite{Ruess2020JES} where the influence of wetting was modulated using liquid and solid electrolytes. As illustrated in Figures~\ref{fig:benchmark}(d) and (h), the formation of internal fracture surfaces increases total electrochemically active area beyond the external boundary. This expansion of active surface area results in a reduced local current density and associated lithium flux under the constraint of a constant total current, effectively decreasing the local charging rate and enhancing the overall (half-)cell charging capacity. The influence of wetting is further evidenced in the lithium concentration distributions under charging stage shown in Figures~\ref{fig:benchmark}(e) and (i). In the absence of wetting, Li-rich region concentrated internally and Li-poor region near the particle boundary are observed (corresponding to time point A in Figure~\ref{fig:benchmark}(b)); in contrast, an additional active surface is introduced along the mid-height fracture path when wetting is considered, resulting in the formation of two distinct Li-rich zones and a remaining Li-poor region near the active surfaces (corresponding to time point A$^{\prime}$ in Figure~\ref{fig:benchmark}(b)).
	
	Similarly, the discharging stage exhibits a consistent trend in both voltage–capacity behavior and lithium concentration distribution. During lithiation, lithium ions enter the particle through the active surfaces and diffuse inward. The mechanical stress generated under lithiation conditions is lower than that during delithiation,\cite{Chen2024JPS} and as a result, no further crack propagation is observed in either case—with or without wetting. Due to the presence of additional active fracture surfaces formed in the charging stage, phase separation behaviour differs notably between the wetting and non-wetting cases. In the non-wetting case, a ring-shaped Li-poor region forms between internal and external Li-rich zones, whereas in the wetting case, two distinct and spatially separated Li-rich regions emerge. Furthermore, it is evident that the particle with wetting exhibits higher Coulombic efficiency during the first charge–discharge cycle (90.4$\%$ for the wetting case vs. 65.9$\%$ for the non-wetting case), along with improved capacity retention. These simulation results—particularly the voltage–capacity profiles—are in strongly qualitative agreement with previously reported experimental findings.\cite{Ruess2020JES}
	
	\subsubsection{Simulation investigations on polycrystalline NCM particle}
	% \threesubsection{First lowest-level subsection}
	
\begin{figure}[htbp]
	\centering
	\includegraphics[width=\textwidth]{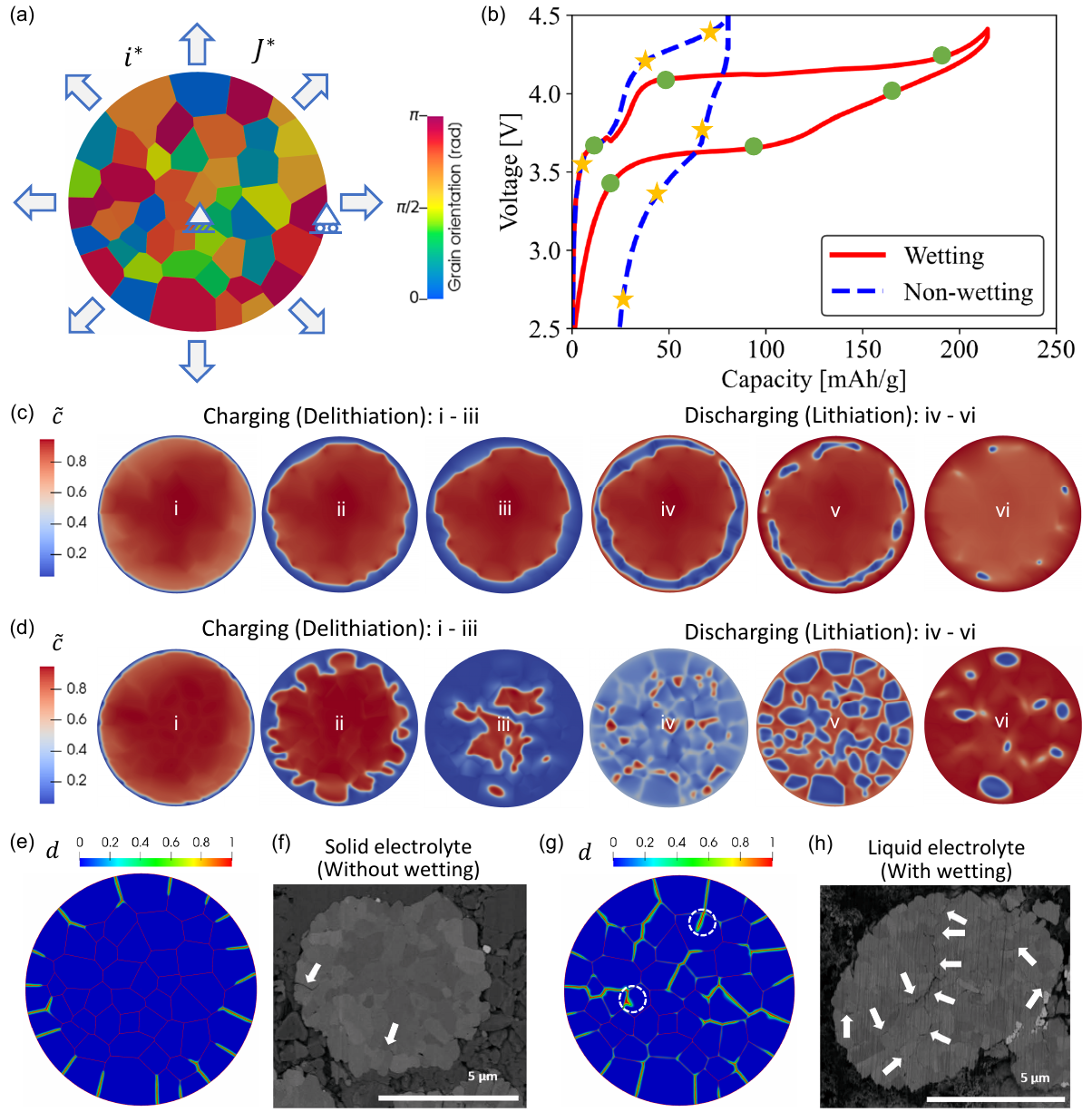}
	\caption{Electro-chemo-mechanical-fracture simulation on polycrystalline NCM particle under wetting and non-wetting regimes. 
	Panel (a) illustrates the simulated polycrystalline microstructure with randomly distributed grain orientations, together with properly defined boundary conditions for multi-physical processes. 
	Panel (b) compares voltage-capacity profiles during the first constant-current charge–discharge cycle for both regimes. 
	Panel (c) shows the lithium concentration maps in the charging and discharging stages in the non-wetting regime, extracted at time points marked in yellow stars in Panel (b). Panels (d) shows the lithium concentration maps in the charging and discharging stages in the wetting regime, corresponding to the time points marked in green circle in Panel (b).
	Panels (e) and (g) show the simulated particle fracture patterns without/with wetting effect, respectively. 
	Panels (f) and (h) show the experimentally observed particle fracture patterns without/with wetting effect, by using solid (f) or liquid (h) electrolyte, respectively. \textcolor{black}{Reproduced from an open-access publication by IOP Publishing (on behalf of The Electrochemical Society), under the terms of the Creative Commons Attribution 4.0 License (CC BY).\cite{Ruess2020JES} %\textcolor{red}{is there significance to the white lines?}
	}}
	\label{fig:polycrystalline}  
	\end{figure}

	\hspace{4mm}
	In this example, we present an electro-chemo-mechanical-fracture simulation of a polycrystalline NCM particle, with a particular focus on investigating the critical influence of the wetting phenomenon and comparing the results with previously published experimental studies.\cite{Ruess2020JES,Trevisanello2021AEM} As illustrated in Figure~\ref{fig:polycrystalline}(a), the particle is subjected to appropriately defined electrical, chemical, and mechanical boundary conditions, consistent with previous simulations. Specifically, a single charge–discharge cycle is applied under a constant current of 1C, starting from an initial lithium concentration of $\tilde{c}_0 = 0.99$, with lower and upper cut-off voltages set to 2.5 V and 4.5 V, respectively. The polycrystalline microstructure is generated using the open-source software Neper,\cite{Neper} and the corresponding finite element mesh is constructed and exported using Gmsh.\cite{Gmsh} Random crystallographic orientations are assigned to grains, as also visualized in Figure~\ref{fig:polycrystalline}(a).
	
	Figure~\ref{fig:polycrystalline}(b) shows the voltage–capacity curves for a constant-current charge–discharge cycle, comparing cases with and without the wetting phenomenon. Figures~\ref{fig:polycrystalline}(c) and (d) illustrate the corresponding evolution of lithium concentration distributions for the non-wetting and wetting cases, respectively. The specific time points at which these concentration maps are extracted are indicated by yellow stars (non-wetting) and green circles (wetting) in Figure~\ref{fig:polycrystalline}(b). During the charging stage, the non-wetting particle exhibits an internal Li-rich region surrounded by an external Li-poor phase, with irregular morphologies (see sub-panels ii and iii in Figure~\ref{fig:polycrystalline}(c)) resulting from the random anisotropy introduced by grain orientations. In contrast, fracture-induced wetting significantly alters the phase morphology. For example, a concave-shaped Li-rich region (see sub-panel ii in Figure~\ref{fig:polycrystalline}(d)) forms adjacent to edge cracks, and internally propagated cracks lead to spatially separated Li-rich regions within the particle (see sub-panel iii in Figure~\ref{fig:polycrystalline}(d)). Overall, the presence of additional active fracture surfaces beyond the external boundary effectively reduces the local charging rate, resulting in an increased accessible capacity under prescribed cut-off voltage, as shown in Figure~\ref{fig:polycrystalline}(b). During the discharging stage, in the non-wetting case, lithium flux enters the particle solely through the external boundary, resulting in the formation of an intermediate Li-poor region between Li-rich zones (see panels iv and v in Figure~\ref{fig:polycrystalline}(c)). In contrast, for the wetting case, both the external boundary and internal fracture surfaces serve as active sites for electrochemical reactions and lithium ingress. This increased active surface area not only effectively lowers the local current density and charging rate but also enhances the equivalent diffusivity and promotes more homogeneous lithium distribution throughout the particle.\cite{Ruess2020JES,Trevisanello2021AEM,Li2023EES} As a result, this leads to improved lithiation uniformity (see panels iv and v of Figure~\ref{fig:polycrystalline}(d)) and contributes to a higher Coulombic efficiency during the first charge–discharge cycle (see Figure~\ref{fig:polycrystalline}(b), 99.4$\%$ for the wetting case vs. 70.7$\%$ for the non-wetting case). The simulated voltage–capacity profiles are in strongly qualitative agreement with experimental observations previously reported in the literature.\cite{Ruess2020JES, Trevisanello2021AEM}
	
	Figures~\ref{fig:polycrystalline}(e) and (g) present the simulated final fracture patterns in NCM particles without and with the consideration of wetting effects, respectively. Corresponding experimental observations obtained via scanning electron microscopy are shown in Figures~\ref{fig:polycrystalline}(f) and (h),\cite{Ruess2020JES} where solid and liquid electrolytes were used to suppress or activate wetting, respectively. The simulation results reveal that wetting significantly enhances fracture evolution, resulting in a higher crack density and more extensive cracking both near the particle surface and within the interior. In the wetting case, both inter-granular cracks along grain boundaries and trans-granular cracks within individual grains (see dashed white circles in Figures~\ref{fig:polycrystalline}(g)) are observed. In contrast, only inter-granular cracks are seen in the non-wetting case. These simulated fracture patterns comparing wetting/non-wetting cases show strong agreement with the experimental findings reported by Ruess et al.\cite{Ruess2020JES} It is worth noting that, based on both simulation results and supporting experimental observations, fracture in cathode particles exposed to liquid electrolytes contributes to enhanced specific charging capacity and Coulombic efficiency after the first cycle. However, chemo-mechanical fracture is widely believed to negatively affect long-term cycling performance due to increased internal resistance, loss of ionic and electronic connectivity, and related degradation mechanisms.\cite{Ruess2020JES,delamination,Janek2019LongTerm,AFM3_design} A detailed investigation of these long-term effects is beyond the scope of the present study.
	
	\section{Conclusion}
	
	%In this study, we utilized advanced experiment techniques, including growth of single-crystal $\alpha$-V$_2$O$_5$ samples, FIB sectioning to prepare lamellae as single-domain model electrodes interfaced with a current collector, chemical (de)lithiation, and high-resolution electron and X-ray imaging to map fracture evolution patterns and electrolyte-wetting-induced heterogeneous lithium distribution. These methods enable a comprehensive investigation of the complex chemo-mechanical coupling between fracture and wetting dynamics in lithium intercalation host materials. Meanwhile, promising simulation tools, specifically, a multi-physical model with capability to simulate concurrent bulk and interface fracture and their coupling with wetting phenomenon, was developed and adopted to consistently capture chemo-mechanical coupling observed in experiments, with an excellent agreement reached regarding heterogeneous lithium concentration maps and complex ultimate failure modes. 
	
	\hspace{4mm}
	In this study, we investigated the intricate interplay between wetting-enhanced diffusion dynamics and chemo-mechanical fracture in lithium-ion battery cathode materials through an integrated experimental and simulation approach. To isolate fundamental coupling mechanisms, advanced experimental techniques and validated simulations were conducted on single-crystal $\alpha$-V$_2$O$_5$, which offers minimal microstructural complexity compared to polycrystalline materials. Electron and X-ray microscopy enabled direct mapping of fracture patterns and wetting-induced heterogeneous lithium distributions following controlled chemical lithiation–delithiation cycles. These experimentally observed behaviors were consistently reproduced by simulations, providing clear evidence of wetting–fracture coupling mechanism. Building on this foundation, the validated modeling framework was extended to multi-physics simulations on polycrystalline NCM particles, where the critical role of wetting in fracture propagation and local/global electrochemical responses was numerically demonstrated, in strong agreement with previously reported experimental findings. 
	
	Evidence from $\alpha$-V$_2$O$_5$ single-crystal studies reveals that the interplay between liquid electrolyte wetting and chemo-mechanical fracture manifests in two fundamental ways. First, wetting at newly formed (bulk/interface) fracture surfaces induces additional surface reactions and (de)lithiation flux, enhancing compositional heterogeneity and global (de)lithiation efficiency. Second, wetting significantly influences fracture dynamics by (i) introducing new fracture modes, such as internal bulk cracking, (ii) accelerating crack propagation, and (iii) modulating fracture directionality in conjunction with crystal orientation (e.g., orthogonal to the preferred diffusion pathway). These coupled mechanisms operate concurrently, synergistically amplifying (de)lithiation capability and fracture behaviours in cathode materials. Multiphysical simulations on polycrystalline NCM particles further reveal the critical role of the wetting phenomenon. Under constant charging current and controlled cut-off voltages, particles exhibiting wetting effects demonstrate enhanced charging capacity and improved Coulombic efficiency. These improvements are attributed to the increased electrochemically active surface area, reduced local lithium flux, and shortened diffusion pathways enabled by wetting-induced internal fracture surfaces. Additionally, wetting leads to more extensive fracture propagation and higher crack density within the particle.
	
	The results of this study carry important implications for the design of electrode materials aimed at enhancing capacity through controlled crack engineering. On one hand, fracture formation facilitates electrolyte infiltration along newly formed surfaces, effectively increasing the electrochemically active area and reducing lithium diffusion paths. This, in turn, promotes faster lithiation/delithiation kinetics and improves capacity utilization. Such insights inform the rational design of electrolyte systems—optimizing parameters such as morphology, particle shape, size, and orientation—to achieve high-rate performance. On the other hand, uncontrolled fracture propagation remains a critical factor in long-term capacity fade, due to structural degradation and loss of electrical and ionic pathways. Thus, a careful balance must be achieved between the short-term benefits of fracture-enhanced kinetics and the long-term detriments associated with mechanical degradation.\\
	%The investigations further illustrate the distinctive utility of single crystals of intercalation hosts as model systems to examine foundational multifield and multiphysics coupling. Besides, the findings of this study have significant implications for both single-crystal and polycrystalline electrodes design with improved capacity and mitigated mechanical degradation. On one hand, chemo-mechanical fracture can induce capacity fade by isolating active material islands \cite{Wang2019,Edge2021}. On the other, crack formation and elastic energy release facilitate electrolyte infiltration along fracture surfaces, reducing diffusion lengths and increasing reactive surface area, thereby enhancing lithiation/delithiation kinetics and capacity utilization. For single-crystal cathodes, the results highlight the advantages of aligning geometries with crystal symmetry to minimize stress gradients and fracture risk. For polycrystalline electrodes, they inform the design of nanostructured 3D architectures that mitigate mechanical degradation by strategically patterning electrolyte infiltration pathways to accommodate stress \cite{Xia2016situ,chen2014hierarchical,Andrews2020Matter,NanostructuredElectrodes,Jiao2023review}. \\
	
	% Experimental section
	% \section{Experimental Section}
	% \threesubsection{First part of experimental section}\\
	% \threesubsection{Second part of experimental section}\\
	
	%%%%%%%%%%%%%%%% Supporting Information
	\noindent
	\medskip
	\textbf{Supporting Information} %\par %Please delete the Suppporting Information statement if it is not applicable. Please supply Supporting Information in another file. Supporting information should not be provided in .tex format
	
	Supporting Information is available from the Wiley Online Library or from the author. \\

	%%%%%%%%%%%%%%%% Conflict of Interest
	\noindent
	\medskip
	\textbf{Conflict of Interest}
	
	The authors declare no conflict of interest.\\

	%%%%%%%%%%%%%%%% Data Availability Statement
	\noindent
	\medskip
	\textbf{Data Availability Statement}
	
	The data that support the findings of this study are available from the
	corresponding author upon reasonable request.\\

	%%%%%%%%%%%%%%%% Acknowledgements
	\noindent
	\medskip
	\textbf{Acknowledgements} \par %delete if not applicable))
	The authors Chen and Xu gratefully acknowledge the computing time granted on the Hessian High-Performance Computer ``Lichtenberg'' (Project-02017, Special-00007). This work has been (partially) funded by the German Research Foundation (DFG) under grant 460684687. \textcolor{black}{Part of the research described in this article was performed at the Advanced Light Source (COSMIC beamline). The Advanced Light Source is supported by the Director, Office of Science, Office of Basic Energy Sciences, of the US Department of Energy (DOE), under contract No. DE-AC02-05CH11231. We thank Dr. David Shapiro for his help with scanning transmission X-ray microscopy/ptychography. Part of the research described in this article was performed at the Canadian Light Source (SM beamline), a national research facility of the University of Saskatchewan, which is supported by the Canada Foundation for Innovation (CFI) , the Natural Sciences and Engineering Research Council of Canada (NSERC), the Canadian Institutes of Health Research (CIHR), the National Research Council Canada, the Canadian Institutes of Health Research, the Government of Saskatchewan, and the University of Saskatchewan. We thank Dr. Jian Wang for his help with scanning transmission X-ray microscopy/ptychography. We acknowledge National Science Foundation (NSF) Awards CMMI2038625 as part of the NSF/DHS/DOT/NIH/USDA-NIFA Cyber-Physical Systems Program and DMR 1809866. Use of the Texas A$\&$M University Materials Characterization Facility (RRID: SCR$\_$022202) is acknowledged. We thank Dr. Sisi Xiang for her help with the transmission electron microscope and focused ion beam.}

	%%%%%%%%%%%%%%%% References
	\noindent
	\medskip
	%\textbf{References}\\
	%\bibliographystyle{unsrtnat}
	%\bibliographystyle{unsrtnat}
	\bibliography{refer}

	\clearpage
	\begin{spacing}{1.0}
		
		{\RaggedRight
			% \title{Deciphering the interplay between wetting and chemo-mechanical fracture in ($\alpha$-V$_2$O$_5$ single-crystal) positive electrode materials}
			\title{The Supporting Information for \\ ``Deciphering the interplay between wetting and chemo-\\mechanical fracture in \textcolor{black}{lithium-ion battery cathode materials"}}
			\maketitle
		}
		
		% Author: Please give full first and last names for authors and include * after the name of all corresponding authors
		
		{\RaggedRight
			\author{Wan-Xin Chen$\ddag$}
			\author{Luis J. Carrillo$\ddag$}
			\author{Arnab Maji}
			\author{Xiang-Long Peng}
			\author{Joseph Handy}\\
			\author{Sarbajit Banerjee*}
			\author{Bai-Xiang Xu*}
		}
		
		% Dedication
		\noindent \dedication{$\ddag$: These authors contributed equally to this work.}
		
		% Affiliations: Please provide adacemic titles (Prof. or Dr.) for all authors where applicable, and include an institutional email address for all corresponding authors
		\begin{affiliations}
			\noindent Wan-Xin Chen, Xiang-Long Peng, Bai-Xiang Xu\\
			Division Mechanics of Functional Materials, Institute of Materials Science, Technischen Universit\"at Darmstadt, Darmstadt 64287, Germany\\
			Email Address: \textcolor{blue}{xu@mfm.tu-darmstadt.de}\\
			
			\noindent Luis J. Carrillo, Arnab Maji, Joseph Handy \\
			Department of Materials Science and Engineering, Texas A$\&$M University, College Station, Texas 77843-3003, United States\\
			Department of Chemistry, Texas A$\&$M University, College Station, Texas 77842-3012, United States \\ 
			
			\noindent Sarbajit Banerjee\\ 
			Department of Materials Science and Engineering, Texas A$\&$M University, College Station, Texas 77843-3003, United States\\
			Department of Chemistry, Texas A$\&$M University, College Station, Texas 77842-3012, United States \\ 
			Laboratory for Inorganic Chemistry, Department of Chemistry and Applied Biosciences, ETH Zurich, Vladimir-Prelog-Weg 2, CH-8093 Z\"urich, Switzerland\\
			Laboratory for Battery Science, PSI Center for Energy and Environmental Sciences, Paul Scherrer Institute, Forschungsstrasse 111, CH-5232 Villigen PSI, Switzerland\\
			Email Address: \textcolor{blue}{sarbajit.banerjee@psi.ch}
		\end{affiliations}
		
	\end{spacing}

\clearpage
\section{Experimental Method}
\subsection{Growth of $\alpha$-V$_2$O$_5$ single crystals}

\hspace{4mm}
Adapted from previous synthesis,\cite{handy2022li} $\delta$-Li$_{0.7}$V$_2$O$_5$ powders were firstly prepared using a solvothermal process. Stoichiometric amounts of LiOH (Sigma Aldrich, 99.6$\%$), V$_2$O$_5$ (Sigma Aldrich, 99.6$\%$), and 86 mL of EtOH were added to a polytetrafluoroethylene-lined stainless-steel autoclave (Parr, 125 mL capacity) and allowed to react for 72 h at 210$^{\circ}$C. The resulting black powder was filtered and allowed to dry overnight. The powder was ground and annealed, placed in an alumina ceramic boat, which in turn was placed in a fused silica tube. The powders were then heated under a flow of Ar gas at 600$^{\circ}$C for 12 h to remove residual moisture (Thermo Fisher Scientific, Lindberg Blue M with UT150 Controller). To obtain large crystals, the resulting powder was ball-milled (Spex mill) using acrylic beads, sealed in a fused silica ampoule under vacuum, then melted at 800$^{\circ}$C. and cooled at a rate of 2$^{\circ}$C$/$h in a programmable furnace (Thermo Scientific, Lindberg Blue M with UT150 controller) to obtain large black lustrous plate-like single crystals.

To obtain $\alpha$-V$_2$O$_5$ single crystals, topochemical deintercalation of $\delta$-Li$_{0.7}$V$_2$O$_5$ was performed by treating a batch of single crystals with 1.5 M equivalents of NOBF$_4$ (Alfa Aesar, 98$\%$) in dry acetonitrile (ca. 0.01 M solution) for 24 h at room temperature.\cite{Santos2022,Santos2022_2} Topochemical de-intercalation of Li-ions is accompanied by a dramatic change in color from lustrous black to yellow/orange. A layer-like habit was observed upon imaging the V$_2$O$_5$ crystals by SEM.

\subsection{FIB etching and lift out}

\hspace{4mm}
Large single crystals with lateral dimensions 100-200 $\upmu$m were fixed to a SEM stub using two-sided carbon tape (Ted Pella, 20 mm width by 20 m length). FIB-SEM imaging and sample preparation were performed using a Tescan LYRA-3 equipped with a Schottky field-emission electron source and fully integrated Canion Ga FIB column. The instrument also contained a five-reservoir gas injection system (GIS) with W, Pt, SiO$_x$, H$_2$O, and XeF$_2$. Lift out was facilitated by a SmarAct 3-Axis piezo nanomanipulator.

Briefly, FIB lift out was done by etching out a flat surface of the single crystal to define a trench with lateral dimensions of 10 $\upmu$m $\times$ 2 $\upmu$m trench using a 12 nA ion beam current. The trench was then polished and thinned to 10 $\upmu$m $\times$ 0.4 $\upmu$m using a 3 nA beam current. Subsequently, a “U-cut” was performed using a 1.3 nA beam current in preparation for lift out. The lamella was then fixed to a W rod (Ted Pella, W with 0.508 mm shank diameter) nanomanipulator, etched with a beam current of 200 pA, lifted out, and attached to a FIB Cu grid (PELCO, single Cu post). Finally, a current of 50 pA was used to thin out the lamellae to approximately 250 nm for STXM analysis.

\subsection{Lithiation and Delithiation of FIB-Sectioned Samples}

\hspace{4mm}
%Lithiation of V$_2$O$_5$ lamella was performed chemically using n-butyllithium (2.5M in hexanes, Sigma, Aldrich) diluted to 0.5M in n-hexane (DriSolv, $\geq$95\%, Sigma Aldrich), submerged for 10 seconds (\textbf{Equation A.1a}).
Topochemical lithiation of $\alpha$-V$_2$O$_5$ lamellae was performed using \textit{n}-butyllithium (2.5M in hexanes, Sigma Aldrich) diluted to 0.5M in $n$-hexane (DriSolv, $\geq$95$\%$, Sigma, Aldrich) by immersing the single post Cu grid with the lamellae attached in a 10 mL solution for 10 s as per:
%\begin{subequations}\label{eq:lithiation}
\begin{align}
	\label{eq:lithiaion}
	\text{V}_2\text{O}_5\left ( \textit{s}\right) + \textit{x}\text{LiC}_4\text{H}_9\left (\textit{hex.}\right ) \quad\longrightarrow \quad \text{Li}_\textit{x}\text{V}_2\text{O}_5\left ( \textit{s}\right ) + \frac{\textit{x}}{2}\text{C}_8\text{H}_{18}\left ( \textit{l}\right )
\end{align}
%\end{subequations}
%
Delithiation was performed using the same apparatus using a 0.5M solution of nitrosonium tetrafluoroborate (98$\%$, Thermo Fisher) dissolved in acetonitrile (DrySolv, $\geq$95$\%$, Sigma, Aldrich). The single post Cu grids with lamellae attached were immersed again in 10 mL for 10 s to deinsert Li-ions as per: 
%
%\begin{subequations}
\begin{align}\label{eq:delithiaion}
	\text{Li}_\textit{x}\text{V}_2\text{O}_5\left (\textit{s} \right ) + \textit{x}\text{NOBF}_4\left (\textit{acetonitrile}\right ) \quad \longrightarrow \quad \text{V}_2\text{O}_5\left (\textit{s}\right ) +\textit{x}\text{Li}\text{BF}_4\left (\textit{s}\right ) + \textit{x}\text{NO}\left (\textit{g}\right )
\end{align}
%\end{subequations}
%
Copious washes were performed with $n$-hexane after lithiation and acetonitrile after delithiation. Briefly, after the first lithiation, 3 washes with 10 mL $n$-hexane were performed by immersing the single Cu grid with the lamellae attached for 30 s followed by immersion in 10 mL acetonitrile for 30 s. After delithiation, 3 washes were performed with 10 mL of acetonitrile by immersing the s Cu grid with the lamellae attached for 30 s, followed by immersion in 10 mL on $n$-hexane for 30 s. 

It's worth noting, repeated cycling led to an increased formation of Li$_2$O through surface conversion reactions accompanying intercalation: V$_2$O$_5$ (s) $+$ 2C$_4$H$_9$Li (solv.) $\longrightarrow$ 2VO$_2$ (s) + Li$_2$O (s) + C$_8$H$_{18}$ (solv.), which is consistent with previous findings.\cite{canepa2017odyssey,Wang2016PRB}

\subsection{Scanning/Transmission Electron Microscopy}

\hspace{4mm}
High-resolution TEM (HRTEM) images were acquired using an FEI Titan Themis 300 S/TEM instrument. The finished lifted-out lamella was mounted directly onto a single-tilt holder for analysis. The Titan Themis is equipped with an extreme field-emission gun (X-FEG) equipped with a monochromator to improve acceleration voltage resolution to 300 KV $\leq$ 0.2 eV and sub-Ångstrom spatial resolution. Furthermore, the TEM also is equipped with a Super-X EDS detector (Thermo Fisher Scientific, FEI) with 4 windowless detectors and an energy resolution at Mn K$\alpha$ with $\pm$128 eV.

\subsection{Scanning Transmission X-ray Microscopy}

\hspace{4mm}
STXM measurements were performed at the 10ID-1 beamline of the Canadian Light Source in Saskatoon, SK, CA and at the coherent scattering and microscopy beamline (COSMIC) of the Advanced Light Source at Lawrence Berkeley National Lab in Berkeley, CA, USA. The 10ID-1 beamline is equipped with an elliptically polarized Apple II type undulator, which provides an intense beam in the 130–2700 eV energy range, whereas the COSMIC beamline has 6 elliptically polarizing inductors, providing an intense beam in the 250–2500 eV energy range. State-of-the-art Fresnel zone plate optics were utilized in conjunction with an order-sorting aperture to achieve a focused X-ray beam. 

\begin{figure}[htbp]
	\centering
	\includegraphics[width=0.9\textwidth]{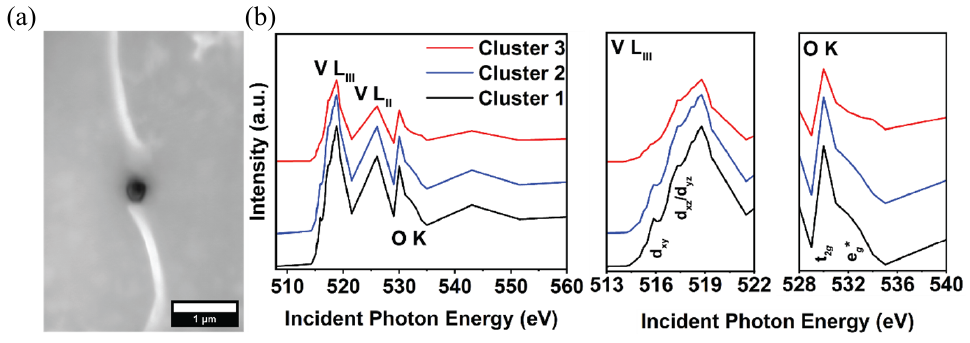}
	\caption{(a). Average STXM image corresponding to Figure 2(l) in the manuscript; (b). Corresponding XANES spectra from the cluster analysis for Figure 2 (l), assignments to V L- and O K-edge spectra are denoted.
	}
	\label{fig:appendix-experiment}  
\end{figure} 

Regions of interest were defined based on the dimensions of the lamellae. Regions of interest were usually found using a single energy point of around 518 (V-L3 absorption) to achieve the highest contrast. Before collection of spectromicroscopy stacks, a line-scan was performed to view average composition and ensure accurate V and O absorption features. Spectromicroscopy stacks were obtained in the range 508—560 eV corresponding to V L- and O K edges. A step size of 0.2 eV was used in regions of spectral interest, whereas spectra in post-edge and pre-edge regions were collected with a step size of 1 eV. Typically, the region of interest is raster-scanned whilst simultaneously recording transmission intensity in a stepwise fashion using a charged-coupled device detector. Image registration and processing were achieved utilizing the cross-correlation analysis “Jacobsen stack analyze” function in the aXis2000 software suite (Version April 22, 2023). An incident spectrum was extracted from the region corresponding to the transmission of silicon nitride substrate to convert the overall transmission measurements to an optical density matrix. Principal component analysis (PCA) and subsequent k-means clustering were implemented utilizing the “PCA$\_$GUI” routine (version 1.1.1) within the aXis2000 suite.\cite{Santos2022,hitchcock2023analysis} The quantity of principal components considered for clustering were chosen based on eigenvalues and their corresponding eigenspectrum/eigenimage representation of the data. Previously published libraries of spectra acquired at different extents of lithiation aid in preliminary interpretation of eigenspectra,\cite{Santos2022} which provide a Bayesian means for the selection of significant components. The averaged spectrum from each cluster was compared to standards published previously to discern Li stoichiometries for each cluster, and subsequently used to obtain phase-specific signatures from which composition maps could be generated by singular value decomposition (vide infra). See Figure~\ref{fig:appendix-experiment} for an example of STXM image and the corresponding XANES spectra.

\section{Simulation Method}

\subsection{Derivation of phase-field fracture model considering wetting effects}

\hspace{4mm}
Here, we provided the thermodynamically consistent derivations of the multi-physical model. The primary unknowns include the molar concentration per unit volume $c (\boldsymbol{x}, t)$, chemical potential $\mu (\boldsymbol{x}, t)$, displacement field $\boldsymbol{u} (\boldsymbol{x}, t)$ (vector quantity in bold) and crack phase-field $d (\boldsymbol{x}, t)$, with $\boldsymbol{x}$ and $t$ labeling the material point coordinate and time, respectively. Under the small-deformation setting, the total strain tensor $\bfepsilon (\boldsymbol{x})$ is obtained via the symmetric part of the displacement gradient tensor, which can be further decomposed into two parts, i.e., elastic $\bfepsilon_{e}$ and chemical $\bfepsilon_{c}$ strain tensors.
\begin{subequations}\label{eq:kinematics}
	\begin{align}
		\label{eq:kinematics-total}
		&
		\bfepsilon = \nabla ^{\text{sym}} \boldsymbol{u} = \dfrac{1}{2} (\nabla \boldsymbol{u} + \nabla^{\rm{T}} \boldsymbol{u}) = 
		\bfepsilon_{e} + \bfepsilon_{c},
		\\
		\label{eq:kinematics-chemical}
		&
		\bfepsilon_{c} = (c - c_0) \boldsymbol{\Omega}, 
		\qquad \boldsymbol{\Omega} = \Omega/3 \boldsymbol{I}.
	\end{align}
\end{subequations}
The elastic part directly contributes to the mechanical stress, while the chemical one measures the material volume change induced by varying concentration within the host material during (dis)charging processes. Here $c_0$ denotes the initial concentration in the active material under stress-free state, $\boldsymbol{\Omega}$ is the chemical deformation (expansion/shrinkage) coefficient tensor.

The chemical flux $\bfJ$ and mechanical stress tensor $\bfsigma$ are governed by the following equations:
\begin{subequations}\label{eq:governing-equations}
	\begin{align}
		\label{eq:governing-equations-chemical}
		&\begin{cases}
			\nabla \cdot \boldsymbol{J} + \dot{c} = Q
			\qquad \quad &
			\; \text{in} \; \varOmega \\
			\boldsymbol{J} \cdot \boldsymbol{n} = J^{\ast} 
			\qquad \quad &
			\; \text{on} \; \partial \varOmega_{J} \\
			c = c^{\ast}
			\qquad \quad &
			\; \text{on} \; \partial \varOmega_{c} \\
		\end{cases}, \\
		\label{eq:governing-equations-mechanical}
		&\begin{cases}
			\nabla \cdot \bfsigma = \boldsymbol{0} 
			\qquad \quad &
			\text{in} \; \varOmega \\
			\bfsigma \cdot \boldsymbol{n} = \boldsymbol{t}^{\ast}
			\qquad \quad &
			\text{on} \; \partial \varOmega_{t} \\
			\bfu = \bfu^{\ast}
			\qquad \quad &
			\text{on} \; \partial \varOmega_{u} \\
		\end{cases}.
	\end{align}
\end{subequations}
Here prescribed chemical flux $J^{\ast}$ and concentration $c^{\ast}$ respectively applied on boundary $\partial \varOmega_{J}$ and $\partial \varOmega_{c}$ serve as chemical boundary conditions, similarly, prescribed traction force $\boldsymbol{t}^{\ast}$ (on $\partial \varOmega_{t}$) and displacement $\boldsymbol{u}^{\ast}$ (on $\partial \varOmega_{u}$) are boundary conditions on the external surface $\partial \varOmega$ in the mechanical sub-problem. The crack induced chemical wetting is considered by constructing a source term $Q$ within the phase-field fracture context, which can be proved to be equivalent to the sharp interface approach.

The constitutive equations for the multi-physical model and phase-field governing equations are derived in terms of the Clausius-Duhem inequality under isothermal condition.\cite{Coleman1967} For the chemo-mechanically coupled system, one can write the total dissipation considering the quantities as following:
\begin{align}\label{eq:energy-dissipation}
	\dot{\mathscr{D}} = 
	-\int_{\partial \varOmega} J^{\ast} \mu \; \mathrm{d} A
	-\int_{\varGamma_{+}} J^{\ast} \mu \; \mathrm{d} A
	-\int_{\varGamma_{-}} J^{\ast} \mu \; \mathrm{d} A 
	+ \int_{\partial \varOmega} \boldsymbol{t}^{\ast} \cdot \dot{\bfu} \; \mathrm{d} A
	- \int_{\varOmega} \dot{\psi} \; \mathrm{d} V 
	\geq 0.
\end{align}
In the above, time derivation is denoted by $\dot{()}$, $\varGamma$ denotes the sharp crack surface with $\pm$ representing its two opposite sides, i.e.,  $\psi$ represents internal potential energy density taking into account the contributions from all processes including chemical, mechanical and fracture sub-problems, i.e., 
\begin{subequations}\label{eq:potential-energy}
	\begin{align}
		\label{eq:potential-energy-total}
		& \psi = \psi_c(c, \nabla c)
		+ \psi_m(\bfepsilon (\bfu), c, d)
		+ \psi_d(\bfx, d, \nabla d),
		\\
		\label{eq:potential-energy-chemical}
		& \psi_c = R T c_{\max } [\tilde{c} \ln \tilde{c}+(1-\tilde{c}) \ln (1-\tilde{c}) + \chi \tilde{c} (1 - \tilde{c})] + \dfrac{1}{2} c_{\max } \kappa | \nabla \tilde{c} |^2,
		\\
		\label{eq:potential-energy-mechanical}
		& \psi_m = \dfrac{1}{2} \bfepsilon_e: \omega (d) \mathbb{E}_0: \bfepsilon_e
		= \dfrac{1}{2} \left( \bfepsilon - \bfepsilon_c \right) 
		: \omega (d) \mathbb{E}_0 : \left( \bfepsilon - \bfepsilon_c \right),
		\\
		\label{eq:potential-energy-damage}
		& \psi_d = G(\bfx) \; \gamma(d , \nabla d) = \dfrac{G(\bfx)}{\pi} \left( \dfrac{2d - d^2}{b} + b |\nabla d|^2 \right).
	\end{align}
\end{subequations}
In above formulations, $R = 8.314$ J/$ (\text{mol} \cdot \text{K})$ is the gas constant, $T = 298.15$ K is the reference temperature, $c_{\max}$ denotes the maximum Li concentration in active material for normalizing Li concentration $\tilde{c} = c/c_{\max}$, $\chi$ is the interaction parameter which controls the phase separation phenomenon, e.g., $\chi > 2$ allows the coexistence of two phases,\cite{Zhao2016CMAME} $\kappa$ is related to the interfacial energy/thickness parameter;\cite{Zhao2016CMAME} $\mathbb{E}_0$ is the anisotropic elasticity stiffness tensor, with elastic constants listed in Table~\ref{table:material-parameters-chemo-mechanical}.\cite{Reed2020PRB,Jachmann2005SSC,Singh2017PCCP} It should be noted that, lithium-concentration-dependent elastic property is ignored in the current simulations, which will not affect the qualitative analysis of the results.\cite{Shahed2021JMPS,Wang2019EA}
%where $\mathbf{1}$ and $\mathbb{I}$ are the unit second- and fourth- order tensors, respectively, $\lambda_0=v_0 E_0/[\left(1-2 v_0\right)\left(1+v_0\right)]$ and $\mu_0=E_0/[2\left(1+v_0\right)]$ are the Lam\'{e} constants of isotropic elasticity, expressed in terms of Young’s modulus $E_0$ and Poisson’s ratio $\nu_0$ of the material. 
The energetic degradation function $\omega (d)$ in the cohesive phase-field fracture model is expressed in terms of crack phase-field $d$ as following,\cite{Wu2017JMPS} 
\begin{align}\label{eq:energy-degradation-function}
	\omega (d) 
	= \dfrac{(1-d)^2}{(1-d)^2 + a_{1} d ( 1- 0.5 d)}, \quad \text{with} \quad
	a_{1} = \dfrac{4 G(\bfx) E_0}{\pi b \sigma_c^2},
\end{align}
where $\sigma_c$ is the failure strength (critical stress) under uni-axial tensile test, $b$ is the length-scale parameter which determines the width of the transition zone between the unbroken and broken region. As mentioned in the manuscript, the interface $\varGamma_{I}$ between the platinum and the active material is diffusively treated with exponential distribution of fracture energy, so that in the whole domain we have the global fracture energy equivalence,\cite{Chen2024JMPS} i.e., 
\begin{align}\label{eq:3D-energy-consistence}
	\int_{\varOmega} G(\bfx) \gamma(d, \nabla d) \; \mathrm{d} V = 
	\int_{\varGamma_i^{\prime}} G(\bfx) \gamma(d, \nabla d) \; \mathrm{d} V + 
	\int_{\varGamma_b^{\prime}} G_b \gamma(d, \nabla d) \; \mathrm{d} V 
	= \int_{\varGamma_i} G_i \; \mathrm{d} A  
	+ \int_{\varGamma_b} G_b \; \mathrm{d} A,
\end{align}
with $\varGamma_i$ and $\varGamma_b$ denoting sharp crack surfaces at the interface and in the bulk, respectively, i.e., $\varGamma = \varGamma_b + \varGamma_i$, $\varGamma_i^{\prime}$ and $\varGamma_b^{\prime}$ representing the phase-field regularized sub-domains for the interface crack and bulk crack, respectively.

By plugging in the introduced free energy density in the above and the governing equations in Eq.~\ref{eq:governing-equations} to the dissipation inequality Eq.~\ref{eq:energy-dissipation}, one obtains,
\begin{equation}\label{eq:energy-dissipation-further}
	\begin{aligned}
		\dot{\mathscr{D}} = 
		\int_{\varOmega} 
		\left( \bfsigma - \dfrac{\partial \psi}{\partial \bfepsilon} \right) 
		: \dot{\bfepsilon} \; \mathrm{d} V
		+ \int_{\varOmega} 
		\mu \dot{c} \; \mathrm{d} V
		- \int_{\varOmega} 
		\left( \dfrac{\partial \psi}{\partial c} \dot{c} 
		+ \dfrac{\partial \psi}{\partial \nabla c} \cdot \nabla \dot{c} \right) 
		\mathrm{d} V
		+ \int_{\varOmega} 
		Q \mu \; \mathrm{d} V
		\\
		- \int_{\varGamma_{+}} J^{\ast} \mu \; \mathrm{d} A
		- \int_{\varGamma_{-}} J^{\ast} \mu \; \mathrm{d} A
		- \int_{\varOmega} 
		\bfJ \cdot \nabla \mu \;
		\mathrm{d} V
		- \int_{\varOmega} 
		\left( \dfrac{\partial \psi}{\partial d} \dot{d} 
		+ \dfrac{\partial \psi}{\partial \nabla d} \cdot \nabla \dot{d} \right) 
		\mathrm{d} V
		\geq 0.
	\end{aligned}
\end{equation}
We can have the following mechanical stress-strain relation from the first term of the inequality,
\begin{align}\label{mechanical-constitutive-law}
	\bfsigma = & \dfrac{\partial \psi}{\partial \bfepsilon}
	= \dfrac{\partial \psi_m}{\partial \bfepsilon}
	= \omega (d) \mathbb{E}_0 : \left( \bfepsilon - \bfepsilon_c \right).
\end{align}
By applying the divergence theorem, the terms in above inequality related to the variation of the chemical concentration can be listed as:
\begin{align}\label{chemical-constitutive-law-1}
	\int_{\varOmega} 
	\left( \mu - \dfrac{\partial \psi_c}{\partial c} 
	- \dfrac{\partial \psi_m}{\partial c}
	+ \kappa \nabla \cdot \nabla \tilde{c}
	\right) \dot{c} 
	\mathrm{d} V
	- \int_{\partial \varOmega} 
	\kappa \bfn \cdot \nabla \tilde{c} \; \dot{c} 
	\mathrm{d} A \geq 0.
\end{align}
According to the Coleman-Noll principle for satisfying the dissipation inequality,\cite{Coleman1967} we have following constitutive laws for chemical potential
\begin{align}\label{chemical-constitutive-law-2}
	\mu = \dfrac{\partial \psi_c}{\partial c}
	+ \dfrac{\partial \psi_m}{\partial c}
	- \kappa \nabla \cdot \nabla \tilde{c}
	= R T \left[ \ln \dfrac{\tilde{c}}{1-\tilde{c}} 
	+ \chi (1-2\tilde{c}) \right] 
	- \bfsigma : \boldsymbol{\Omega}
	% \dfrac{\partial \psi_m}{\partial c}
	- \kappa \nabla \cdot \nabla \tilde{c},
\end{align}
together with the natural boundary condition satisfied automatically
\begin{align}\label{chemical-constitutive-law-boundary}
	\bfn \cdot \nabla \tilde{c} = 0.
\end{align}
The terms in the inequality related to the variation of the crack phase-field yields:
\begin{align}
	\begin{cases}
		% case1
		-\omega'(d) \bar{Y}
		- \dfrac{G(\bfx)}{\pi b} (2-2d)
		+ \dfrac{2b}{\pi} G(\bfx) \nabla \cdot \nabla d = 0,
		\quad \text{when } \dot{d} > 0 \vspace{1mm} 
		\quad \quad &
		\text{in} \; \varOmega\\
		% case2
		-\omega'(d) \bar{Y}
		- \dfrac{G(\bfx)}{\pi b} (2-2d)
		+ \dfrac{2b}{\pi} G(\bfx) \nabla \cdot \nabla d < 0,
		\quad \text{when } \dot{d} = 0 \vspace{1mm} 
		\quad \quad &
		\text{in} \; \varOmega \\
		% surface BC
		\nabla d \cdot \boldsymbol{n} = 0
		\quad \quad &
		\text{on} \; \partial \varOmega
	\end{cases}, 
\end{align}
by assuming the maximum dissipation principle and applying the Lagrange multipliers under Karush–Kuhn–Tucker (KKT) constraints. The effective crack driving force is modified to consider to tension-dominant fracture behaviors,\cite{Wu2017JMPS,Wu2020CMAME} i.e.,
\begin{align}
	\label{eq:crack-driving-force-simple}
	\bar{Y} = \frac{ \bar{\sigma}_{\text{eq}}^2 }{2 \bar{E}_0}, 
	\qquad \text{with} \quad
	\bar{\sigma}_{\text{eq}} = \max( \left\langle \bar{\sigma}_1 \right\rangle, \sigma_c),
\end{align}
where $\bar{\sigma}_1$ is the maximum principle value of the effective stress tensor $\bar{\bfsigma} = \mathbb{E}_0 : \bfepsilon_e$, $\bar{E}_0$ is the elongation modulus. Regarding the fracture induced wetting chemical flux, following terms considering contributions from both interface and bulk cracks are expressed, i.e.,    
\begin{equation}
	\begin{aligned}
		\int_{\varGamma_{+}} J^{\ast} \mu \; \mathrm{d} A + \int_{\varGamma_{-}} J^{\ast} \mu \; \mathrm{d} A 
		= \int_{\varGamma} 2J^{\ast} \mu \; \mathrm{d} A
		= \int_{\varGamma_i} 2J^{\ast} \mu \; \mathrm{d} A + \int_{\varGamma_b} 2J^{\ast} \mu \; \mathrm{d} A \\
		\approx \int_{\varGamma_i^{\prime}} 2 \dfrac{G(\bfx)}{G_i} \gamma(d, \nabla d) J^{\ast} \mu \; \mathrm{d} V + 
		\int_{\varGamma_b^{\prime}} 2 \dfrac{G(\bfx)}{G_b} \gamma(d, \nabla d) J^{\ast} \mu \; \mathrm{d} V,
	\end{aligned}
\end{equation}
besides, the term related to crack induced source term in Eq.~\ref{eq:energy-dissipation-further} can be written as
\begin{align}
	\int_{\varOmega} Q \mu \; \mathrm{d} V
	= \int_{\varGamma_i^{\prime}} Q \mu \; \mathrm{d} V + 
	\int_{\varGamma_b^{\prime}} Q \mu \; \mathrm{d} V,
\end{align}
thus, the source term can be defined as
\begin{align}\label{eq:source-term}
	Q =
	\begin{cases}
		2 \dfrac{G(\bfx)}{G_i} \gamma(d, \nabla d) J^{\ast}
		\quad \text{in the diffusive interface region for potential interface crack} \; \varGamma_i^{\prime} \vspace{2mm} \\
		2 \dfrac{G(\bfx)}{G_b} \gamma(d, \nabla d) J^{\ast}
		= 2 \gamma(d, \nabla d) J^{\ast}
		\quad \text{\cite{Miehe2016IJNME, Zhao2016CMAME, Emilio2023CMAME} in the bulk region for potential bulk crack} \; \varGamma_b^{\prime}
	\end{cases}
\end{align}
On the basis of the properly defined source term for considering the wetting effect, the dissipation inequality yields the last term, we can assume the linear relation between the chemical flux and the chemical potential gradient to ensure to be non-negative,\cite{Chen2024JMPS} i.e, 
\begin{equation}
	\begin{aligned}
		&\bfJ = - \boldsymbol{M} (c, d) \cdot \nabla \mu
		= - \omega (d) \dfrac{c (1-\tilde{c})}{RT} \boldsymbol{D} \cdot \nabla \mu \\
		\quad \Longrightarrow \quad
		&- \int_{\varOmega} \bfJ \cdot \nabla \mu \; \mathrm{d} V
		= \int_{\varOmega} \omega (d) \dfrac{c (1-\tilde{c})}{RT} \nabla \mu \cdot \boldsymbol{D} \cdot \nabla \mu \; \mathrm{d} V \geq 0.
	\end{aligned}	
\end{equation}
In above formulations, $\boldsymbol{M} = (1-d)^2 c (1-\tilde{c})/(RT) \boldsymbol{D}$ is concentration-dependent mobility tensor, which is degraded by $(1-d)^2$ in terms of crack phase-field;\cite{Shahed2021JMPS,Chen2024JPS} $\boldsymbol{D} = D_{ij} \boldsymbol{e}_{i} \otimes \boldsymbol{e}_{j}$ denotes the diffusivity tensor with $D_{ij}$ representing its components. In the simulation, diffusivity at the $b$ direction is assumed to be larger than that at the $a$ direction,\cite{Xu2021ACS,Ma2013JPD} see Table~\ref{table:material-parameters-chemo-mechanical} for the specific values.

In the simulations on $\alpha$-V$_2$O$_5$ single-crystal specimens under chemical (de)lithiation, constant flux $J^{*}$ is applied on the cathode material's external boundary and fracture surface; while for polycrystalline NCM particle where electro-chemo-mechanical-fracture simulations are performed, Butler-Volmer (BV) kinetics is introduced to determine the local current density and corresponding lithium flux,\cite{Newman2021Book} i.e., local current density $i_{\text{BV}}$ can be expressed as
\begin{align}\label{eq:BV-current-density}
	i_{\text{BV}} = i_0(\tilde{c}) \left[ 
	\exp \left( \dfrac{F\eta}{2RT} \right) - \exp\left( -\dfrac{F\eta}{2RT} \right)
	\right].
\end{align}
In the above, $F = 96485.3321$ C$/$mol is the Faraday constant, $i_0$ is the lithium-concentration-dependent exchange current density expressed as following,\cite{Zhao2019JMPS}
\begin{align}\label{eq:BV-exchange-current-density}
	i_0 = F k_0 c_{\max} \sqrt{\tilde{c} (1 - \tilde{c})} \left( \dfrac{c_l}{c_{l\_\text{ref}}} \right),
\end{align}
where $k_0$ is the rate constant for the cathodic reaction, $c_l$ and $c_{l\_\text{ref}}$ are two constants denoting respectively the Li concentration in the electrolyte and its corresponding reference value. Besides, $\eta$ in \label{eq:BV-current-density} is the overpotential, which can be expressed as
\begin{align}\label{eq:BV-overpotential}
	\eta = \phi_c - \phi_e - \phi_{eq}(\tilde{c}),
\end{align}
where $\phi_c$ and $\phi_e$ respectively denotes the potential of the cathode particle and the electrolyte. In the modeling, the former needs to be resolved spatially, while the latter is assumed to be constant (0) and uniform in the particle domain. $\phi_{eq}$ represents the equilibrium potential for Li reaction in the active material,\cite{Zhao2019JMPS} which is dependent on the lithium concentration and expressed as following
\begin{align}\label{eq:BV-equilibrium-potential}
	\phi_{eq} = 4.4 - 2.8\tilde{c} + 8.2\tilde{c}^2 + 9.8\tilde{c}^3 -214.5\tilde{c}^4 + 777.8\tilde{c}^5 - 1290.6\tilde{c}^6 + 1034.4\tilde{c}^7 - 324.2\tilde{c}^8,
\end{align}
other options measured from experiments could also be adopted.\cite{Danner2016JPS,Colclasure2019JES,Marker2019CM}

For the electrochemical boundary condition in the simulation, local current density defined by the BV formulation serves as the electrical boundary current, and the corresponding chemical flux can be determined by using the Faraday constant,
\begin{align}\label{eq:electrochemical-BC}
	i^{\ast} = i_{\text{BV}}, \qquad J^{\ast} = i^{\ast} / F.
\end{align}
In the present study, the Galvanostatic (dis)charging condition is utilized, i.e., the total current demand $I$ at all electrochemically active surfaces (including external boundary $\partial \varOmega$ and fracture surfaces $\varGamma$) is constant during the charging process, 
\begin{align}\label{eq:total-current}
	I = \int_{\partial \varOmega + \varGamma} i^{\ast} \; \mathrm{~d} A.
\end{align}
$I$ characterizes the quantity of electricity carried by the lithium-ions inserting into/extracting from the electrode surface within an unit time, which remains a constant for the entire (dis)charging process. However, it's worth noting the current density at the particle surface $i^{\ast}$ is allowed to vary locally, e.g., resulting from diverse grain orientations and damage levels within the heterogeneous system. In this case, instead of using the uniformly distributed current density/chemical flux at the electrode surface,\cite{Klinsmann2015JES,Bai2021IJSS} we convert the constant total current constraint into a volumetric integral, i.e., 
\begin{align}
	I = \int_{\partial \varOmega + \varGamma} \bfn \cdot \bfi \; \mathrm{~d} A
	= \int_{\varOmega} \nabla \cdot \bfi \; \mathrm{~d} V.
\end{align}
We have
\begin{align}
	\int_{\varOmega} \left( \nabla \cdot \bfi - \dfrac{I}{V} \right) 
	\; \mathrm{~d} V = 0, 
	\quad \text{with} \quad
	V = \int_{\varOmega} \; \mathrm{~d} V,
\end{align}
with $V$ denoting the volume of the cathode particle. As can be seen, the surface integral constraint in Eq.~\ref{eq:total-current} can be implemented with an equivalent source term, so that the electrical governing equation under constant charging current reads
\begin{align}\label{eq:governing-equations-electrical-new}
	\nabla \cdot \bfi - \dfrac{I}{V} = 0 \quad \text{in} \; \varOmega,
	\quad \boldsymbol{i} \cdot \boldsymbol{n} = i^{\ast} = i_{\text{BV}} 
	\quad \text{on} \; \partial \varOmega \; \text{and} \; \varGamma.
\end{align}
It's worth noting that the local current density $i^{\ast}$ at the sharp fracture surface $\varGamma$ --- whether interface $\varGamma_i$ or bulk $\varGamma_b$ cracks, should be represented in a diffusive manner through an equivalent source term. This is formulated based on the phase-field damage variable $d$ within the corresponding smeared sub-domains $\varGamma_i^{\prime}$ and $\varGamma_b^{\prime}$, analogous in form to the chemical flux source term described in Eq.~\ref{eq:source-term}.

\subsection{Simulation setup and adopted parameters}
\hspace{4mm}
To capture the experimental observations on $\alpha$-V$_2$O$_5$ single crystals with high fidelity, we consider a heterogeneous domain $\varOmega$ consisting of the active material $\varOmega_c$ and the surrounding platinum frame $\varOmega_p$, with the in-between sharp interface $\varGamma_I$ diffusively treated, see Figure 1(b) in the paper for an illustration. For the assignment of fracture energy in the computational domain, constant value $G(\boldsymbol{x}) = G_b$ is assigned for the bulk phase, while exponential interpolation $G(\xi)$ is utilized at the diffusive interface region, with effective interface fracture energy $\tilde{G}_i$ determined according to the (global) interface fracture energy equivalence. Details on solving $\tilde{G}_i$ and implementing property interpolation can refer to our previous publication.\cite{Chen2024JMPS} It should be noted, 
%that the aforementioned formulas apply specifically to the active material $\alpha$-V$_2$O$_5$ $\varOmega_c$; for the surrounding platinum $\varOmega_p$, 
the chemical process in the platinum $\varOmega_p$ is neglected, whereas the mechanical and fracture processes are retained to simulate the deformation and interface delamination in the structure. The developed multi-physics model is implemented in the MOOSE platform for finite-element simulations.\cite{moose} All parameters adopted in the simulations are listed in Table~\ref{table:material-parameters-chemo-mechanical}.
\begin{table}[htbp]
	\centering
	\caption{Material parameters adopted in the simulations. \cite{Bai2021IJSS,Reed2020PRB,Jachmann2005SSC,Singh2017PCCP,McGraw1999EA,Santos2022,de2018striping,Xu2021ACS}}
	\label{table:material-parameters-chemo-mechanical}
	\begin{tabular}{ m{0pt} m{7cm}<{\centering} m{3cm}<{\centering} m{2.5cm}<{\centering} }
		\toprule[1pt]
		\rule{0pt}{10pt} &
		Material parameters & Value & Unit \\
		\hline
		%
		%\hline
		\rule{0pt}{10pt} 
		& Elastic parameters $C_{11}=C_{22}$ 
		& 230 
		& [GPa] \\
		% 
		%\hline
		\rule{0pt}{10pt} 
		& Elastic parameter $C_{12}=C_{21}$ 
		& 116 
		& [GPa] \\
		% 
		%\hline
		\rule{0pt}{10pt} 
		& Elastic parameter $C_{66}$ 
		& 43
		& [GPa] \\
		% 
		%\hline
		\rule{0pt}{10pt} 
		& Bulk tensile strength $\sigma_{c,b}$ 
		& 450
		& [MPa] \\
		% 
		%\hline
		\rule{0pt}{10pt} 
		& Interface tensile strength $\sigma_{c,i}$ 
		& 300
		& [MPa] \\
		%\hline
		\rule{0pt}{10pt} 
		& Bulk fracture energy $G_{b}$ 
		& 2.5 
		& [N/m] \\
		% 
		%\hline
		\rule{0pt}{10pt} 
		& Interface fracture energy $G_{i}$ 
		& 1.0 
		& [N/m] \\
		%\hline
		\rule{0pt}{10pt} 
		& Maximum Lithium concentration $c_{\max}$ 
		& 38320
		& [mol/m$^3$] \\
		% 
		%\hline
		\rule{0pt}{10pt} 
		& Partial molar volume $\Omega_a = 2 \Omega_b$ 
		& 3.497 $\times$ 10$^{-6}$
		& [m$^3$/mol] \\
		% 
		%\hline
		\rule{0pt}{10pt} 
		& Diffusivity $D_b = 10 D_a$ 
		& 7 $\times$ 10$^{-15}$
		& [m$^2$/s] \\
		% 
		% \hline
		\rule{0pt}{10pt} 
		& Phase-field length-scale $b$ 
		& 0.06
		& [$\upmu$m] \\
		% 
		% \hline
		\rule{0pt}{10pt} 
		& Interface length-scale $L$ 
		& 0.06
		% & [$\upmu$m] \\
		& [$\upmu$m] \\
		%
		% \hline
		\rule{0pt}{10pt} 
		& Applied surface flux $J^{\ast}$ 
		& 4.58 $\times$ 10$^{-5}$
		% & [$\upmu$m] \\
		& [mol/(m$^2$s)] \\
		\bottomrule[1pt]
	\end{tabular}
\end{table}

In the simulations of polycrystalline NCM particles under electrochemical (dis)charging conditions governed by Butler–Volmer kinetics, additional parameters are required. These include the phase parameter $\chi = 2.5$, interface parameter $\kappa = 2.5 \times 10^{-10}$ J·m$^2$/mol, reaction rate constant $k_0 = 2 \times 10^{-11}$ m/s, lithium concentration in the electrolyte $c_l = 1500$ mol/m$^3$, and its reference value $c_{l_\text{ref}} = 1$ mol/m$^3$.\cite{Zhao2016CMAME,Xu2019JMPS,Smith2006JPS} For detailed implementation of the fracture energy interpolation strategy within the polycrystalline domain, readers are referred to our prior work.\cite{Chen2024JMPS}

\section{Benchmark example for the wetting model}

\begin{figure}[htbp]
	\centering
	\includegraphics[width=1.0\textwidth]{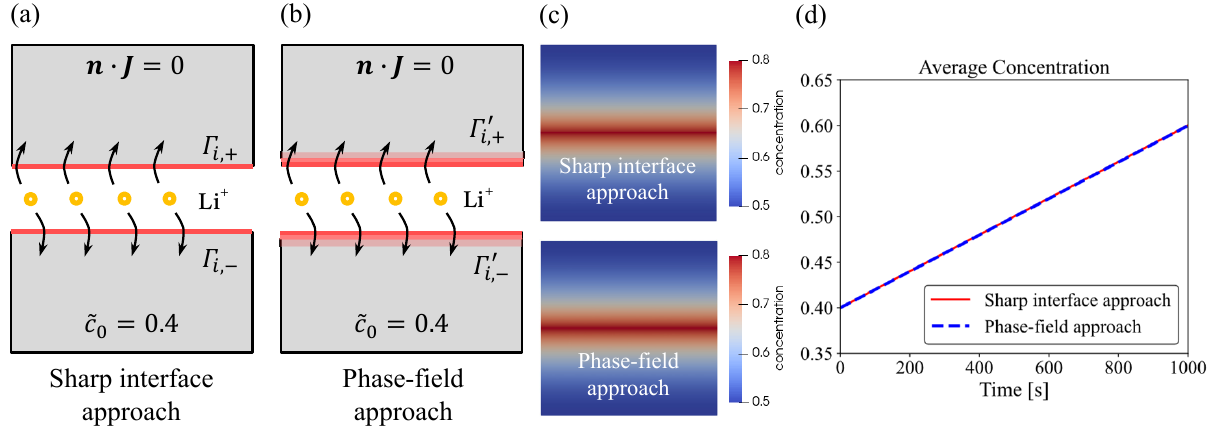}
	\caption{The comparison of utilizing the sharp interface and phase-field approaches to model the interface fracture induced wetting flux in the specimen. Panels (a) and (b) respectively illustrates two approaches in the modeling, i.e., $\int_{\varGamma_{i,\pm}} J^{\ast} \delta c \; \mathrm{d} A$ for the sharp one and $\int_{\varGamma_{i,\pm}^{\prime}} 2 \dfrac{G(\bfx)}{G_i} \gamma(d, \nabla d) J^{\ast} \delta c \; \mathrm{d} V$ for the phase-field approach. Panel (c) compares the concentration profile results obtained by two approaches at the moment $t=1000$ s. Panel (d) compares the evolution of the average concentration in the domain.
	}
	\label{fig:appendix-sharp-phase-field}  
\end{figure}

\hspace{4mm}
In this part, a benchmark example for validating the proposed wetting model is presented, specifically, the chemical flux induced by the predefined interface crack $\varGamma_i$ is simulated using two approaches, i.e., the sharp interface approach and the continuum phase-field framework with introduced source term $Q$ in Eq.~\ref{eq:source-term}. As shown in Figure ~\ref{fig:appendix-sharp-phase-field}, a specimen consisting of the upper and lower parts are connected with a middle interface, and a predefined crack is assumed to be located exactly at the interface, so that electrolyte infiltration on the surface can introduce additional chemical flux within the sample. Other parts of the surfaces are assumed to be chemically isolated for simplification.
%the surface becomes wetted and liquid chemical flux can enter the specimen under lithiation condition. 

In the sharp interface approach, the chemical flux $J^{\ast}$ on two sides of the interface crack $\varGamma_{i,\pm}$ can be directly modelled with the interface element, as shown in Figure~\ref{fig:appendix-sharp-phase-field}, under the lithiation condition, the average concentration in the domain rises with the time, and local high concentration at the interface crack can be observed in the profile result. The phase-field approach regularizes the interface crack $\varGamma_{i,\pm}^{\prime}$ after prescribing the crack phase-field $d = 1$ at the interface $\varGamma_i$; subsequently, the chemical process is activated via the crack-phase-field-dependent source term $Q$ in Eq.~\ref{eq:source-term}. As can be observed, concentration rises in the domain globally, and the evolution of average concentration quantitatively agrees with that by the sharp interface approach, which proves the validity of the phase-field framework for considering the wetting effect and its equivalence to the standard sharp interface approach. \\

\end{document}